\documentclass[12pt]{article} 
\usepackage[hyperfootnotes=false]{hyperref}
     \usepackage{amsmath}
      \usepackage{amssymb}
  \usepackage{graphicx}
  \usepackage{xcolor}
  \newlength{\abstractwidth}
 \usepackage{mciteplus}

  \newcommand{\be}{\begin{equation}}
  \newcommand{\bea}{\begin{eqnarray}}
  \newcommand{\eea}{\end{eqnarray}}
  \newcommand{\beq}{\begin{equation}}
  \newcommand{\ee}{\end{equation}}
  \newcommand{\eeq}{\end{equation}}

  \newcommand{\half}{{1\over 2}}

  \newcommand{\tl}{\hat{t}}
  \newcommand{\ul}{\hat{u}}

\def\la{\label}

\def\32{{3 \over 2 } }

  \def\ba{\begin{eqnarray}}
  \def\ea{\end{eqnarray}}

 \def\simleq{\; \raise0.3ex\hbox{$<$\kern-0.75em
      \raise-1.1ex\hbox{$\sim$}}\; }
 \def\simgeq{\; \raise0.3ex\hbox{$>$\kern-0.75em
      \raise-1.1ex\hbox{$\sim$}}\; }

\def\taul{{\hat \tau }}
\def\tl{{\hat t}}
\def\ul{{\hat u}}
\def\Sch{{\rm Sch}}
\def\nref#1{(\ref{#1})}

\begin{document}

\begin{titlepage}
 % \rightline{}
  \bigskip

  \bigskip\bigskip

  \bigskip

\begin{center}
%\centerline
{\Large \bf { Conformal symmetry and its breaking in two \\ \bigskip
 dimensional Nearly Anti-de-Sitter space }}
 \bigskip
%\centerline
{\Large \bf { }} 
    \bigskip
\bigskip
\end{center}

  \begin{center}

 \bf {Juan Maldacena$^1$, Douglas Stanford$^1$  and Zhenbin Yang$^2$   }
  \bigskip \rm
\bigskip
 
   $^1$Institute for Advanced Study,  Princeton, NJ 08540, USA  \\
\rm
 \bigskip
 $^2$Jadwin Hall, Princeton University,  Princeton, NJ 08540, USA

 % \bf {Write authors  }
  \bigskip \rm
\bigskip
 
 %   Institute for Advanced Study,  Princeton, NJ 08540, USA  \\
\rm

\bigskip
\bigskip

% \vspace{2cm}
  \end{center}

 \bigskip\bigskip
  \begin{abstract}

 We study a two dimensional dilaton gravity system, recently examined
  by Almheiri and Polchinski,  which 
 describes near extremal black holes, or more generally, nearly $AdS_2$ spacetimes. 
 The asymptotic symmetries of $AdS_2$ are all the  time reparametrizations of the boundary. 
 These symmetries are spontaneously broken by the $AdS_2$ geometry and they are explicitly 
 broken by the small  deformation  away from $AdS_2$. 
 This pattern of spontaneous plus explicit symmetry breaking governs the gravitational backreaction of the system.
   It determines several gravitational properties such as the linear in temperature
 dependence of the near extremal entropy as well as the gravitational corrections to correlation functions. 
 These corrections include the ones determining the growth of out of time order correlators that is indicative 
 of chaos.   
 These gravitational aspects  can be described in terms of a Schwarzian derivative effective action for a 
 reparametrization.

 \medskip
  \noindent
  \end{abstract}
\bigskip \bigskip \bigskip

  \end{titlepage}

  %  \starttext \baselineskip=17.63pt \setcounter{footnote}{0}
   \tableofcontents

 % \sc

\section{Introduction}

In some respects $AdS_2$ is a bit harder to understand than its higher dimensional siblings. 
The main reason is that pure gravity in $AdS_2$ is inconsistent with the existence of finite 
energy excitations
above the $AdS_2$ vacuum  \cite{Fiola:1994ir,Maldacena:1998uz,Almheiri:2014cka}. Nevertheless, there is a sense in which 
{\it nearly } $AdS_2$ gravity is well defined. In  nearly $AdS_2$, or $NAdS_2$, one keeps the leading 
order correction away from $AdS_2$. These corrections have a universal form and a very nice analysis 
of this system was given by Almheiri and Polchinski \cite{Almheiri:2014cka}.

The leading order gravitational effects can be described by a particular dilaton gravity 
system first studied by Jackiw \cite{Jackiw:1984je} and Teitelboim \cite{Teitelboim:1983ux}. 
Dilaton gravity theories in two dimensions have no propagating degrees of freedom. 
In this case, the effective action  is determined by the symmetries of the problem. This symmetry is 
a spontaneously and explicitly broken reparametrization symmetry. 
In the perfect $AdS_2$ limit the system develops a reparametrization symmetry, $t \to \tilde t(t)$. 
This arises as the asymptotic symmetry of $AdS_2$ \cite{Hotta:1998iq,Cadoni:1999ja,NavarroSalas:1999up}
Alternatively, we expect it 
from the tracelessness of the stress tensor of a putative boundary theory. Since that stress tensor has only one
component, its tracelessness due to the scaling symmetry implies that it is zero.
 Thus in the perfect $AdS_2$ 
limit the system has a full reparametrization symmetry. Under such circumstances one would naively 
expect the full system to be topological. In some sense it is, the pure $AdS_2$ limit can only describe
the ground states and their  entropy  \cite{Strominger:1996sh, Sen:2008yk}.

In fact,  this reparametrization symmetry is spontaneously broken. It is spontaneously 
broken because it is only an asymptotic symmetry.  Only an SL(2) subgroup is unbroken by the 
geometry of $AdS_2$. A spontaneously broken symmetry has  associated zero modes. If we think in terms of the 
Euclidean path integral, there is  an infinite number of non-compact 
 zero modes: all the fourier modes of
the reparametrization symmetry. 

However, the symmetry is also explicitly broken because we have decided to keep the leading order
correction away from the conformal limit. This correction is given by the simplest local action 
that is invariant under a global SL(2) symmetry, the Schwarzian derivative of the reparameterization. This action 
was  originally found by Kitaev in his analysis of certain quantum mechanical 
fermion models with emergent reparametrization symmetry \cite{KitaevTalks}. Those models have a similar 
realization of these symmetries (see also \cite{Anninos:2016szt}).   

The effective coefficient in front of this Schwarzian action becomes small at 
low energies or low temperatures. 
For this reason, gravitational corrections, though formally suppressed by the Newton constant,
 lead to important  effects in the IR.  

This action is responsible for the form of the near extremal black hole entropy, which is linear in the temperature. 
It also leads to  important corrections for correlation functions, such as the connected four point function. 
In particular, it controls  the chaos related, exponentially growing,  corrections to out of time order correlators.  

Here we show how this Schwarzian action emerges in detail and we study some of its consequences as
well as some details regarding its physical interpretation.  A similar perspective, but for $AdS_3$, was discussed in \cite{Turiaci:2016cvo}.

While this paper was in preparation similar results were presented in \cite{Jensen:2016pah}. We have also learnt that 
\cite{Engelsoy:2016xyb} have been following similar ideas. 

This paper is organized as follows. 
In section two we review the asympototic symmetries of $AdS_2$,  and we explain how they give rise to a family of geometries once we cutoff
the space. These give rise to an infinite family of zero modes. 
In section three we include the leading perturbation away from the $AdS_2$ limit which corresponds to the near extremal limit of black holes.
 This gives rise to a non-zero action for the above geometries. 
The leading order action describes a near extremal entropy that is linear in the temperature. 
In section four we add matter and show how to couple it to the nearly zero modes described above. Using this coupling we perform some
perturbative computations. These include the gravitational corrections to the four point function, including the out of time order correlator, 
a one loop correction to the free energy and a gravitational correction to the two point function. 
In section five we discuss some Lorentzian aspects of this action. We discuss its SL(2) symmetries and why they 
cure a problem involving higher derivatives.  In section six we discuss how to resum all corrections in the chaos regime to describe the crossover
at the scrambling time. 
 
\section{ Pure $AdS_2$  }

\subsection{Coordinate systems }

\begin{figure}[ht]
\begin{center}
\includegraphics[scale = 0.4]{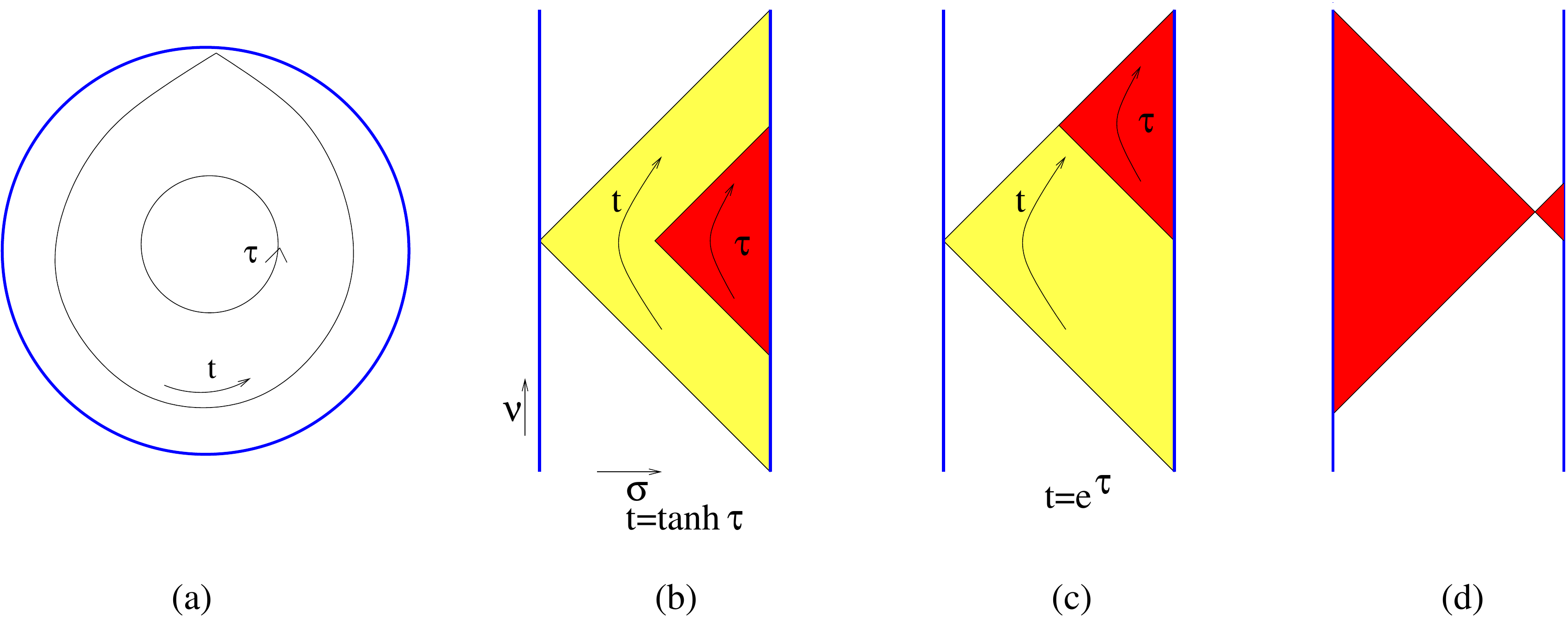}
\caption{ (a) Hyperbolic space or Euclidean $AdS_2$. The orbits of $\tau$ translations look like circles. Orbits of $t$ are curves that
touch the boundary at $t=\pm\infty$. (b) Lorentzian $AdS_2$. The $\nu, \sigma$ coordinates cover the whole strip. The $\tl, z$ coordinates 
describe the Poincare patch denoted here in yellow. The red region is covered by the $\taul , \rho$ coordinates. There are different choices for how
to place the $\taul , \rho$ region that are generated by SL(2) isometries. In (b) and (c) we show two choices and give the relation between the
Poincare time $\tl $ and the   $\taul $ at the boundary of the space. In (d) we show a generic pair of Rindler wedges.   }\label{CoordFig}
\end{center}
\end{figure}

On $AdS_2$ (with unit radius) it is convenient to use the following coordinate systems 
\bea \la{EuCoord}
{\rm Euclidean: } ~~~ &&ds^2 = { dt^2 + dz^2 \over z^2 } ~,~~~~~ ds^2 = d\rho^2 + \sinh^2 \rho d\tau^2 
\\
{\rm Lorentzian: } ~~~ &&ds^2 = { -d \tl ^2 + dz^2 \over z^2 } ~,~~~~~ ds^2 = d\rho^2 - \sinh^2 \rho d\taul^2 ~,~~~~~ \notag
%\cr
 %&& 
  ds^2 =  {- d \nu^2 + d \sigma^2 \over \sin^2 \sigma }  
 \\ \la{Coord}
 {\rm Embedding: } ~~~ &&  -Y_{-1}^2 - Y_0^2 + Y_1^2 =-1 ~,~~~~ds^2 = - dY_{-1}^2 - d Y_0^2 + dY_1^2 
\eea
With Euclidean signature  both coordinate choices 
 cover all of hyperbolic space. In Lorenzian signature they cover different regions of the global space. The hatted versions of the times are Lorentzian e.g. $t = i\hat{t}$. 
The causal structure of the global space is displayed clearly in the $\nu,\sigma$ coordinates. The $\tl ,z$ and $\taul, \rho$ coordinates 
cover different patches as seen in figure \ref{CoordFig}. 
The $\taul, \rho$ coordinates can be viewed as describing the exterior of a finite temperature black hole. 
We can also view them as Rindler coordinates of $AdS_2$.  Note that the finite temperature and zero 
temperature solutions are just different coordinate patches of the same space. 
\subsection{Symmetries and a family of solutions } 

Let us imagine that we have a spacetime that is exactly $AdS_2$, with a finite Newton constant. 
Then the gravitational action is
\be \la{EinsteinTwo}
 I = - \frac{\phi_0}{16\pi G} \left[ \int d^2x \sqrt{g} R  + 2 \int K \right]  + I_m[ g, \chi ]
\ee
where $I_m$ is the matter action  and $\chi$ are the matter fields. Here $\phi_0$ is a constant, which sets the entropy $S_0 = \frac{ \phi_0}{4G}$. 
In two dimensions $G$ is dimensionless. 

We now want to imagine  a situation where this spacetime arises as a low energy limit of a well defined
UV theory. For this purpose we imagine that we cut off the spacetime. 
The UV theory has some time coordinate $u$. Thoughout the paper, we denote the time in the boundary theory by $u$. 
Let us say that the (Euclidean) $AdS_2$ spacetime has the metric in \nref{EuCoord}. 
We want to  cut off the space along a trajectory given by $(t(u),z(u))$.
 We expect to    fix the proper length of the boundary curve 
     \be \la{bdyequ}
      g|_{\rm bdy} = {1 \over \epsilon^2 } ~,~~~~~~~~~  { 1 \over \epsilon^2 }=g_{uu} = {{t'}^{\, 2} + {z'}^{\, 2} \over z^2}%\sqrt{ {t'}^{\, 2} + {z'}^{\, 2} \over z^2}
         \longrightarrow  z = \epsilon t' + O(\epsilon^3)  
      \ee 
      where primes are $u$ derivatives. 
  Note that,  given an arbitrary $t(u)$,   we can  
choose   $z(u) = \epsilon t'(u) $ in order to obey the above equations.  
      Since all other fields are constant on the $AdS_2$ vacuum, 
      when we set the boundary conditions for all the fields to be such constants, we will obey all other boundary conditions. 
Therefore we find that we have a   family of  solutions to the problem, given by $t(u)$.
 
 \begin{figure}[h]
\begin{center}
\includegraphics[scale=.4]{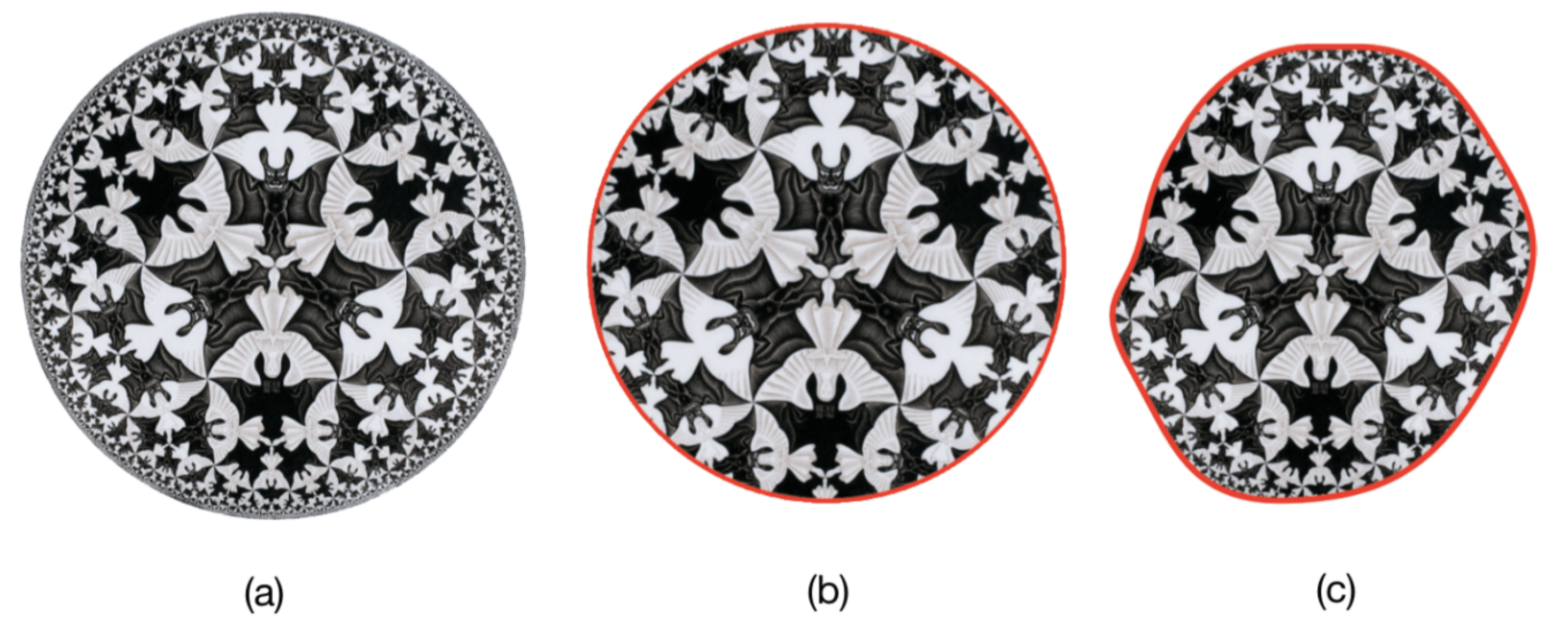}
\caption{In (a) we see the full $AdS_2$ space. In (b) we cut it off at the location of a boundary curve. In (c) we choose a more general boundary 
curve. The full geometry of the cutout space does depend on the choice of the boundary curve. On the other hand, the geometry of this cutout 
region remains the same if we displace it or rotate it by an SL(2) transformation of the original $AdS_2$ space.   }
\label{boundarycut}
\end{center}
\end{figure}

Let us clarify in what sense these are different solutions. The main point is that we are cutting out a
region of $AdS_2$, with different shapes that depend on   the function $t(u)$, see figure \ref{boundarycut}. 
Though the interior $AdS_2$ space is locally the same, the full cutout shape does depend on $t(u)$. 
For example, correlation 
functions of matter fields will depend on the shape chosen by the function $t(u)$. 
Note however, that overall translations or rotations of the whole shape in hyperbolic space do not change the physics. 
These are described by the action of an $SL(2)$ symmetry group   on $AdS_2$. It acts by sending   
\be
 \la{SLTwo} 
 t(u) \rightarrow  
 \tilde  t(u) = { a t(u) + b \over c t(u) + d } ~,~~~~{\rm  with}  ~~
ad -c b =1
\ee
We see that $t(u)$ or $\tilde t(u)$   produce exactly the same cutout shape. 
 Therefore the full  set of different interior geometries is given by the set of all functions $t(u)$  up to the 
above $SL(2)$ transformations. (Or modded out by these SL(2) transformations \nref{SLTwo}). 
 
It is worth noting that we can also look at the asymptotic symmetries of $AdS_2$.  They are generated by   reparametrizations of
the asymptotic form 
\be \la{Atrans}
 \zeta^t = \varepsilon(t) , ~~~~\zeta^z = z \varepsilon'(t) 
 \ee
These will map one boundary curve into another. In fact, \nref{Atrans} sends the curve $t(u)=u$ to  
 $t(u) = u + \varepsilon(u)$.

If we insert these geometries into the action \nref{EinsteinTwo} the Gauss-Bonnet theorem implies that we always get the same action, namely
the extremal entropy. Thus we have a set of exact zero modes parametrized by $t(u)$ (up to the SL(2) identification \nref{SLTwo}).  

Notice that,  near the boundary,  the geometries are indistinguishable, we need to go through the bulk in 
order to distinguish them. In fact, this is the realization of the full reparametrization 
symmetry that we expect in this
problem. In other words, we expect that SL(2) is enhanced to a full Virasoro like symmetry, which in this case, are just 
the reparametrization symmetries. 
  However, the reparametrization symmetry is spontaneously broken by $AdS_2$.  It is broken to 
$SL(2,R)$. The zero modes are characterized by the functions $t(u)$. These can be viewed as 
Goldstone bosons. Except that here we consider them in the Euclidean problem. We can call these
zero modes ``boundary gravitons''. They are similar to the ones that appear in three dimensions.
An important difference with the three dimensional case is that, here,  these modes have precisely zero action in the confromal limit, 
there is no local conformal invariant action we can write down for them.

\section{ $NAdS_2$, or nearly $AdS_2$  spacetimes }

 The pure $AdS_2$ gravity theory discussed above is not consistent with any configuration with non-zero energy, since
the variation of the metric imposes that the stress tensor of matter is identically zero. The Einstein term 
is topological and does not contribute to the equation of motion for the metric. If one is only interested
in understanding the ground state entropy this can be enough \cite{Strominger:1996sh, Sen:2008yk}. 

In order to obtain a reasonable gravity theory it is important to consider a nearly $AdS_2$ geometry. In other words, we need to 
keep track of the leading effects that break the conformal symmetry. 
This is a configuration  that still remembers that the conformal symmetry is slightly broken. 
A model that correctly captures a large number of  situations where $AdS_2$ arises from 
a higher dimensional system (or from some otherwise well defined UV theory) is the following \cite{Almheiri:2014cka}
       \be \label{TwoDAction} 
       I = -\frac{\phi_0}{16\pi G} \left[  \int \sqrt{g} R + 2 \int_{bdy} K \right]   -\frac{1}{16\pi G}\left[ \int d^2x  \phi  \sqrt{g} (R+2 )  + 2\int_{bdy}  \phi_b K \right ]  + I_M[ g, \chi ] + \cdots 
       \ee             
       Here we imagine that $\phi_0 \gg \phi$ and the dots denote higher order terms in $\phi$. We will neglect all such higher order terms here. 
       $\phi_b$ is the boundary value of $\phi$. 
        If $AdS_2$ is arising from the near horizon geometry of an near
       extremal black hole, then $\phi_0 + \phi$ is the area of the two sphere, and $\phi_0$ is the area of the extremal black hole, with $\phi$ denoting the deviations
       from this extremal value. The middle term in the action is the Jackiw Teitelboim two dimensional gravity theory \cite{Teitelboim:1983ux,Jackiw:1984je}.
        The first term is purely topological and its only role is to give the      extremal entropy. We have included the extrinsic curvature terms at the boundary to make the metric variational problem well defined.
         From now on, we will ignore the dots in \nref{TwoDAction}. Since the first term in the action is topological we will also ignore it. 
       
      A thorough analysis of this model was presented in an article by Almheiri and Polchinski \cite{Almheiri:2014cka}.
       Here we simply   emphasize how the pattern of breaking of the reparametrization symmetry determines many aspects of the theory. 
Now, let us analyze the equations of motion of the Jackiw Teitelboim theory 
  \be \la{TJ}
I_{JT} =-  \frac{1}{16\pi G}\left[  \int d^2x  \phi  \sqrt{g} (R+2 )  +2  \int_{bdy}  \phi_b K \right ].
    \ee    
The equations of motion for $\phi $ imply that the metric has constant negative curvature or 
is $AdS_2$. This is also the case if we include the matter term in \nref{TwoDAction} since it 
is independent of the dilaton $\phi$. 
The equations of motion for the metric are 
\be \la{PhiEqu}
T^\phi_{\mu \nu } \equiv 
\frac{1}{8\pi G}( \nabla_\mu \nabla_\nu \phi - g_{\mu \nu} \nabla^2 \phi  + g_{\mu \nu } \phi)  =0  
     \ee
Due to the Bianchi identity, this implies that $T^\phi_{\mu \nu}$ is automatically 
conserved. It turns out that the  general solution is
\be \la{zexpr}
\phi = { \alpha + \gamma  t + \delta (t^2 + z^2)  \over z } = Z. Y 
\ee
where we also rewrote the expression in embedding coordinates \nref{Coord}, where $Z$ is an arbitrary vector.\footnote{More precisely, in \nref{zexpr}  we use the Euclidean version of the embedding coordinates.}  
 
The solution breaks the $SL(2) $ isometries to $U(1)$. In fact, the vector $\zeta^\mu = \epsilon^{\mu \nu} 
\partial_\nu \phi $ is aways a Killing vector of the metric thanks to the equations \nref{PhiEqu} \cite{Mann:1992yv}. 
Thus, the combined dilaton gravity theory always preserves this isometry. 
 
Since $\phi$ is diverging near the boundary, we now have a new dimensionful coupling constant which 
is the strength of that divergence. In other words, beyond the condition \nref{bdyequ} we also need
to impose the condition 
\be \label{phiequ}
\phi_b  = \phi |_{\rm bdy} = { \phi_r(u) \over \epsilon } 
\ee
where $\phi_r(u)$ is an arbitary dimension $-1$ coupling. The $r$ stands for ``renormalized,'' in the sense that it remains finite in the 
$\epsilon \to 0 $ limit. 
For generality we have chosen it to depend
on $u$, but we could also choose it to be independent of $u$.  When we choose it to be constant we will denote it as $\bar \phi_r$. 

When we  embed this into  in a full higher dimensional picture,  we have in mind situations where $\phi_b\propto 1/\epsilon $ is large, but 
$\phi_b \ll \phi_0$ so that we are still in the near extremal region.\footnote{
 This type of expansion is somewhat analogous to the slow roll expansion for inflationary
universes. } In other words, we cut off the space before $\phi$ becomes too large. Note that the notion of ``too large'' is really external to the 
theory \nref{TJ}.  

Now, once we impose both \nref{bdyequ} and \nref{phiequ} we determine completely the shape of
the curve or reparametrization $t(u)$. It is simply given by computing $z(u)$ from \nref{bdyequ}, substituting in \nref{zexpr} and then using \nref{phiequ} to obtain 
\be \la{condicu}
{ \alpha  + \gamma t(u) + \delta \,   t(u)^2 \over t'(u) } = \phi_r(u).
\ee

It is interesting that this condition can also be obtained from an effective action for $t(u)$.
A simple way to obtain the effective action is the following. 
Starting from \nref{TJ} we impose the equation of motion for $\phi$ which implies that
we have an $AdS_2$ space. Inserting that into the action \nref{TJ} we find that the first term vanishes and 
we only get the boundary term,  which involves the boundary value of $\phi $ \nref{phiequ},  
\be
I_{TJ } \to- \frac{1}{8\pi G} \int { d u \over \epsilon } {\phi_r(u) \over \epsilon }  K 
\ee
where we also used that the induced metric is given by $du/\epsilon$, \nref{bdyequ}. 
The extrinsic curvature is given by 
\bea
 K &=&  {t' ( t'^{\,2} + z'^{\,2} + z z'' )- z z' t''  \over ( t'^{\,2} + z'^{\, 2} )^{ 3\over 2 } } = 
1 + \epsilon^2 {\rm Sch}(t,u)  ~,~~~~~~~
\cr
{\rm Sch}(t,u)&\equiv& - { 1 \over 2}  { t''^{\,2} \over t'^{\,2} } + \left({ t''\over t'} \right)' 
\eea 
Inserting this into $I_{TJ}$ we get 
\be \la{SchAction}
I = -\frac{1}{8\pi G}\int du \phi_r(u) \Sch(t,u) 
\ee
We see that the zero modes get an action detemined by the Schwarzian. Here $\phi_r(u)$ is an external coupling and $t(u)$ is
the field variable. 

It is interesting to contemplate why we obtained this. We expect that the breaking of conformal 
symmetry should be local along the boundary, and proportional to $\phi_r(u)$. In addition, we
expect to obtain a local action which involves the Pseudo-Nambu Goldstone modes. Since these
are specified by $t(u)$ up to global $SL(2)$ transformations, we conclude that the simplest term is
the Schwarzian action, which is  indeed SL(2) invariant;  $\Sch(t,u) = \Sch( { a t + b \over c t + d } , u ) $. 

Finally, it easy to check that by varying \nref{SchAction} with respect to $t(u)$ we obtain the equation 
\be \la{tequ}
 \left[ { 1 \over t'} \left(  {( t' \phi_r )' \over t' }   \right)' \right]' =0  
 % ~~~\longrightarrow ~~ \bar \phi_r { [ \Sch(t,u) ]' \over t' }  =0
\ee
 which can be easily integrated to \nref{condicu}, where $\alpha, ~\gamma,~\delta$ are integration
constants.\footnote{A fourth integration constant arises by integrating \nref{condicu}.}  
 Thus we see that the action \nref{SchAction}, which is defined purely on the boundary, 
captures the same information as the bulk expression for the dilaton $\phi$. Notice that
this also implies that the equations of motion of the action \nref{SchAction} are equivalent to 
imposing the equations of motion that result from varying the metric, which were not imposed in 
deriving \nref{SchAction}. The time dependence of $\phi_r(u)$ allows us to pick an arbitrary $t(u)$ as 
the saddle point geometry.  On the other hand, we can also remove it by picking a new time coordinate
via $d \tilde u = \bar \phi_r  d u/ \phi_r(u)  $. When $\phi_r(u)$ is constant \nref{tequ} becomes 
$ \bar \phi_r { [ \Sch(t,u) ]' \over t' }  =0$.

The Schwarzian action summarizes many gravitational effects of the model. As we have 
explained, it follows from the symmetries of the problem and its applicability can go beyond 
systems that are described by a local gravity theory. In fact, this Schwarzian action was 
introduced, for these reasons, by Kitaev in his analysis of certain interacting fermion models 
\cite{KitaevTalks} (see \cite{Maldacena:2016hyu} for a description).

\subsection{The near extremal entropy} 

It is convenient to make a change of field variable in the Schwarzian action from $t$ to $\tau$ of the form 
\be
 t = \tan{ \tau \over 2 }  \la{FinT}.
 \ee
 We can then use the general transformation rule for the Schwarzian to find 
 \bea 
 I &=& - C \int du \, \Sch(t,u) =- C \int du \left[ \Sch(\tau,u) + {\tau' }^{\, 2} \Sch(t,\tau) \right] \notag
\\
 &=& - C \int du \left[  \Sch(\tau,u) + { 1\over 2} {\tau'}^{\, 2} \right] ~,~~~~~~C \equiv { \bar \phi_r \over 8 \pi G}.  
  \la{SchT}
 \eea
We could have derived this form of the action by starting with $AdS_2$ in terms of the coordinates 
$ds^2 = d\rho^2 + \sinh^2 \rho d\tau^2 $ and viewing the boundary as parametrized by $\tau(u)$, with 
$\rho(u)$ determined by the analog of \nref{bdyequ}. 

This is an interesting action, whose solutions are $\tau = { 2 \pi \over \beta } u $  (up to SL(2) transformations).
Note that $\tau \sim \tau + 2 \pi $. 
 For these solutions,  only the term involving 
$\tau'^{\, 2 }$ in \nref{SchT} is important.  On such solutions
the action gives 
\be \la{SchS}
\log Z = - I = 2 \pi^2 { C \over \beta } = 2 \pi^2 C T  
% \frac{\pi}{4G}{ \bar \phi_r  \over \beta} 
\ee
which leads to a near extremal entropy $S = S_0 + 4\pi^2C T$ which is linear in the temperature. 
Note that $4 \pi^2 C T$ is also the specific heat. 
 This  
linear in $T$ behavior is a simple consequence of the reparametrization symmetry and its breaking. 

 This gives us only the near extremal entropy. The extremal entropy, $S_0$, can be obtained by adding a purely 
 topological term to the above action of the form 
 \be 
 - I_{\rm top} = { \phi_0 \over 8 \pi G_N}  \int d u ~ \tau'. 
 \ee
 
 It might seem unusual that we reproduce the entropy from a classical action. This is familiar from the bulk 
 point of view, but it seems unusual to reproduce it from a boundary-looking action. However, this is common 
 in discussions of hydrodynamics. In that case, the free energy is reproduced from a classical action. 
 Here the crucial feature is that the solution depends on the temperature through the condition that 
 $\tau$ winds once as we go from $u=0$ to $u =\beta$. 

  We could wonder whether we should consider  solutions where $\tau$ winds $n$ times, 
 $\tau = n { 2\pi \over \beta } u $.  It appears that this
 effective action makes sense only for the case with winding number one.\footnote{ For $n=0$ the $\tau'$ terms in the numerator are a problem. 
 For $n>1$ the small fluctuations around the solution have negative modes.}
 
 Note that this is not a microscopic derivation of the entropy. This is simply phrasing the computation of
 the entropy as a consequence of a symmetry. We have not given an explicit description of the black hole microstates. 
  If one had a microscopic system which displayed 
 this symmetry breaking pattern, then we would microscopically  explain  the form of the entropy. 
 
It is also possible to compute the ADM energy of the system. This is given in terms of the boundary values of the 
fields. In this case,  we get \cite{Almheiri:2014cka}
\be \la{ADM}
M = { 1 \over 8\pi G} { 1\over \epsilon } [ \phi_b -\partial_n \phi ] =  { \bar \phi_r  \over 8\pi  G}  \Sch(t,u)= C \, \Sch(t,u) =-C\, \Sch(\tl,\ul)
\ee
The second expression is giving the mass terms of the Schwarzian action. This can be obtained by either solving the equations for $\phi$ or
by deriving the conserved quantity associated $u$ translations for \nref{SchAction}. It is valid in the absence of boundary sources for massive fields. The general formula is given in (\ref{genphiS}).

\section{ Adding matter} 
\la{AddMat} 

We can now add matter as in \nref{TwoDAction}. 
Since $\phi$ does not appear in the matter action, the metric is still fixed to the $AdS_2$ metric by the $\phi$ equation of motion, and the 
matter fields move on this  fixed $AdS_2$ geometry. 
The gravitational backreaction is completely contained in the equations obeyed by the dilaton, which are simply \nref{PhiEqu} but with the 
matter stress tensor in the right hand side\footnote{This structure is similar to other models of dilaton gravity where the metric is forced to be flat, 
instead of $AdS_2$, 
see \cite{Strominger:1994tn} for a review.}.
 These are three equations for a single variable $\phi$, but the conservation of the matter 
stress tensor implies that
the equations are consistent. 
The boundary is located by finding the  curve where $\phi = \phi_b$.
 This can be done by first solving for $\phi$ in the bulk as described above and
then finding the trajectory of the boundary curve. 
Alternatively, one can show that the final equation for the trajectory is given by an equation where we 
 add a new term in the right hand side of \nref{tequ}. For the case of massless matter fields we obtain 
\be \la{MatEq}
C  { (\Sch(t,u))' \over t' } = - t' T_{t z}
\ee
%JM: I THINK WE NEEDED THIS EXTRA MINUS SIGN,AT LEAST IN LORENTZIAN SIGNATURE,  PLEASE CHECK. \ZY{[I think in Euclidean there is no minus sign while in Lorentzian there is coming from $H=-C\, Sch$]}
 %\JM{ I think that in Lorentzian signature $T_{\hat t , z} < 0$ means that there is energy going into the space, and that should give $\partial_{\ul } M > 0$. }\ZY{That's right.}
A simple derivation is obtained by equating the change in energy \nref{ADM} to the flux of energy, $- t'^2 T_{tz}$, into the space.
%\ZY{[I think here is no minus sign, z is pointing inwards]}
  A factor of $t'$ comes from 
 redshifting the energy from $t$ to $u$ time and another factor from going from energy per unit $t$ to energy per unit $u$ \footnote{In Lorentzian signature we get a minus sign in \nref{MatEq} from the minus in 
\nref{ADM}.}. 
 Solving  \nref{MatEq} we find $t'(u)$ and solve directly for the trajectory of the boundary curve. The correction to (\ref{MatEq}) when we have sources for massive fields is given in appendix \ref{MassiveFields}.

For correlation function  computations it is useful to calculate the effective action as a function of the boundary conditions for the matter fields, 
$\chi_r(u)$, which   can be functions of the boundary time. 

It  is convenient to  solve first an auxiliary problem, which consists of finding the effective action for the matter
fields in $AdS$, with boundary conditions $\tilde \chi_r(t)$ which are functions of the $AdS_2$ boundary time. 
For a free field in $AdS_2$ this is simple to compute and we obtain 
%
%Since the metric that appears in the matter action is simply the $AdS$ metric, we can easily compute
%the effective action after integrating out the massive fields $\chi$ with some boundary condition. 
%We find that 
%\be 
%- I_{eff} = \int dt dt' { \chi_b(t) \chi_b(t')  (z z')^{\Delta -1} \over |t -t'|^{ 2 \Delta } }
%\ee

%\be 
%- I_{eff} = C\int dt dt' { \chi_b(t) \chi_b(t')  (z z')^{\Delta -1} \over |t -t'|^{ 2 \Delta } }; \quad
% C=\frac{\Delta \Gamma(\Delta)}{\sqrt{\pi} \Gamma(\Delta-\frac{1}{2})}
%\text{for $\Delta=1$},b=\frac{1}{\pi}
%\ee
\be \la{CorPoi}
- I_{eff} = D \int dt dt' { \tilde \chi_r(t) \tilde \chi_r(t')    \over |t -t'|^{ 2 \Delta } }
\ee
where $D$ is a constant.\footnote{$D = \frac{(\Delta - \frac{1}{2})\Gamma(\Delta)}{\sqrt{\pi}\Gamma(\Delta-\frac{1}{2})}$, or $D =1/2\pi$ for $\Delta =1$.}
Once we specify the trajectory of the actual boundary curve via $t(u)$ we can transform this to the desired  boundary conditions
\be \la{ChBdyC}
\chi \sim      z^{1-\Delta} \tilde  \chi_r(t)  = \epsilon^{1 -\Delta }  {t'}^{ 1- \Delta} \tilde  \chi_r(t) = \epsilon^{ 1 -\Delta} \chi_r(u) 
\longrightarrow \chi_r(u) = [ t'(u) ]^{ 1 -\Delta} \tilde \chi_r(t(u) ).
  \ee
  The first expression defines $\tilde \chi_r(t)$,  then we used the expression for $z$ from \nref{bdyequ}, and finally we compared it to the 
  expression for $\chi$ that defines $\chi_r(u)$. Using \nref{ChBdyC}, we rewrite \nref{CorPoi} as 
\be \la{sourct}
-I_{eff} = D \int du du'  \left[  { t'(u) t'(u')  \over [t(u)-t(u')]^2 }\right]^{\Delta }  \chi_{r}(u) \chi_r(u' ) .
%~,~~~~ \chi_r(u) \equiv   
%\epsilon^{\Delta -1} \chi_b(t(u) ) 
\ee 

Though we did this for the two point function, the same is true for any $n$ point function. If we had a self interacting matter theory in $AdS_2$ and
we computed the $AdS_2$ $n$ point correlation function, then the correct physical one in $NAdS_2$, after coupling to gravity, would 
be obtained by writing them using $t(u)$, and rescaling by a factor of $t'(u_i)^{\Delta_i}$ at the insertion of each operator.
\footnote{
We can say that if $Z_M[\tilde \chi_r(t), z(t)]$ is the partition function of the pure matter theory with boundary conditions at $z = z(t)$, then 
the one in the theory coupled to dilaton gravity is $Z_{\rm Dressed}[\chi_r(u), \epsilon] = Z_M[ {t'}^{\Delta -1 } \chi(u) , z(t(u))] $, where 
$z =\epsilon t' + \cdots $.}  
Even if we had free fields in $AdS_2$ this coupling to gravity makes them interact with each other. What is remarkable is the simplicity of this coupling. 

In this way we have found a coupling between $t(u)$ and the matter action. We see that the coupling
proceeds by a reparametrization of the original two point function. 
The full correlation functions are obtained by integrating over $t(u)$, after we add the Schwarzian action 
  \nref{SchAction}. 
  These are the same formulas derived in \cite{Almheiri:2014cka}. The classical equations for $t(u)$ that follow from the variation of the 
Schwarzian action, \nref{SchAction}, plus \nref{sourct} are the same as the ones in \cite{Almheiri:2014cka}.

 \subsection{Perturbative expansion of the Schwarzian action } 
  
  Since the Schwarzian action is of order $1/G$ we can evaluate its effects using perturbation theory around a solution. 
 To avoid carrying unnecessary factors of the temperature we set $\beta = 2\pi$. The factors of temperature can be 
reinstated by dimensional analysis. 
We then set 
\be \la{Repar}
\tau =    u  + \varepsilon (  u )   
\ee
 %t is most useful to do this at finite temperature since it will be connected to the chaos computation. 
%At finite temperature it is useful to think of the reparametrization as 
%\be \la{Repar}
%t(u) = \tan { \pi (u + \epsilon(u) ) \over \beta }  
%\ee
%which leads to the following term that is quadratic in $\epsilon$ 
%\be
%I_\epsilon^2 = \int du  ( \epsilon'')^2 - (\epsilon')^2 ~,~~~~~~~~{\rm for} ~~ \beta = 2 \pi 
%\ee
in \nref{SchT} and expand to second order in $\varepsilon$ to  obtain 
\be \la{SchTe}
I_\epsilon ={ C \over 2 } \int du  [  { \varepsilon''}^{\, 2} - { \varepsilon' }^{\, 2} ]~,~~~~~~~~{\rm for} ~~ \beta = 2 \pi 
\ee
We would like to compute the propagator for this action. A problem is that the action has three zero modes, going like 
$\epsilon = 1 , ~ e^{ i u } ,~ e^{ -i u}$. These zero modes arise from SL(2) transformations of the background solution $\tau = u$. 
Recall that these SL(2) transformations  did not generate new geometries. Therefore we should not be integrating over them in the first place, since the 
integral  over $\varepsilon$ is only over distinct geometries. This is equivalent to viewing the SL(2) symmetry as a gauge symmetry, so that we can 
gauge fix those three zero modes to zero and invert the propagator. The answer is
%\be \la{propag} 
%\langle \epsilon(\tau) \epsilon(0) \rangle = { 1 \over \phi_r }  FILL 
%\ee
\be \la{propag} 
\langle \epsilon(u) \epsilon(0) \rangle = {1 \over 2 \pi C  }  \left[ -{(|u|-\pi)^2\over 2} +(|u|-\pi)\sin |u| +a + b \cos u 
%  1+{\pi^2\over 6}+{5\over 2}\cos u 
\right]
\ee
The last two terms are proportional to SL(2) zero modes and cancel in any gauge invariant computation.\footnote{ A direct inversion of the operator
 gives $a = 1 + \pi^2/6$ and $b = 5/2$ \cite{Maldacena:2016hyu}.}
    Note that the propagator is formally suppressed by $G$, but it is enhanced as $\bar \phi_r $ becomes small. 
  We will now use this propagator for some computations. The effective coupling is $\beta/C$ which is the same as the inverse of the
  near extremal entropy.\footnote{This is the reason that there is trouble with naive black hole thermodynamics at $C/\beta \sim 1$ \cite{Fiola:1994ir}. }

  \subsection{Gravitational contributions to the four point function } 
  
Suppose that we have operators $V$, $W$, which are 
 dual to two different fields which are free  in $AdS_2$ before coupling to gravity. The gravitational contribution to the four point function can be computed as follows. (Some four point functions were also considered in 
\cite{Almheiri:2014cka}. 
 These steps are identical to the ones discussed in \cite{Maldacena:2016hyu}, since the 
effective action is the same.)
We  start from the factorized expression for the four point function,
$\langle V(t_1) V(t_2) W(t_3) W(t_4) \rangle = { 1 \over t_{12}^{2 \Delta } } { 1 \over { t_{34}^{ 2 \Delta } } }
$.     
 We then  insert the reparametrizations \nref{FinT} and \nref{Repar} into \nref{sourct}  and expand to linear order in $\varepsilon$ to obtain 
\be \la{Bilo}
{ 1 \over t_{12}^{2 \Delta } } \longrightarrow {\cal B}(u_1,u_2) { \Delta \over \left[2 \sin {u_{12} \over 2 } \right]^{2\Delta } } ~,~~~~~
{\cal B}(u_1,u_2)\equiv \left[ \varepsilon'(u_1)  + \varepsilon'(u_2) - { \varepsilon(u_1) - \varepsilon(u_2)  \over \tan { u_{12} \over 2 } } \right]. 
\ee
We make a similar replacement for $t_{34}^{-2\Delta}$, and then contract the factors of $\varepsilon$ using the propagator \nref{propag}. This gives the $O(1/C)=O(G)$ contribution to the four point function.
Note that the bilocal operator ${\cal B}$ is  SL(2) invariant.\footnote{ 
Indeed $a$ and $b$ in \nref{propag} disappear if we consider $\langle \epsilon(u) {\cal B}(u_1,u_2) 
\rangle $.}
 %We use \nref{Bilo} for each of the factors  to obtain an expression for the two point function as the expectation
%product of two bilocals. We then use    use the propagator \nref{propag} to find the expression for the four point function. 
The final expression depends on the relative ordering of the four points. When  $u_4 < u_3 < u_2 < u_1 $ 
we obtain  the factorized expression  
%\be
%\langle V_1 V_2 W_3 W_4 \rangle = { 1 \over \phi_r } 
%\left( - 2  + { u_{12} \over \tan{ u_{12} \over 2}  } \right) \left( -2 + { t_{34} \over \tan { u_{34} \over 2 } }  \right)  { 1 \over \sin^{2\Delta } {u_{12} \over 2 } } { 1 \over \sin^{2\Delta } {u_{34} \over 2 } }
%\ee
\be \la{Separated}
 { \langle V_1 V_2 W_3 W_4 \rangle_{\rm grav}  \over \langle V_1 V_2 \rangle \langle W_3 W_4 \rangle }  =\Delta^2 \langle {\cal B}(u_1,u_2) 
  {\cal B}(u_3,u_4) \rangle =  {  \Delta^2 \over 2 \pi C  } 
\left( - 2  + { u_{12} \over \tan{ u_{12} \over 2}  } \right) \left( -2 + { u_{34} \over \tan { u_{34} \over 2 } }  \right)   
\ee
As discussed in \cite{Maldacena:2016hyu}, this expression can be viewed as arising from energy fluctuations. Each two point function generates an energy fluctuation which then 
affects the other. Since energy is conserved, the result does not depend on the relative distance between the pair of points. In other words, we can 
think of it as 
\be \la{Eflu}
{ \langle V_1 V_2 W_3 W_4 \rangle }_{\rm grav}  = \partial_M  \langle V_1 V_2 \rangle  \partial_M \langle W_3 W_4 \rangle  { 1 \over - \partial^2 _M S(M)} 
=  \partial_\beta  \langle V_1 V_2 \rangle  \partial_\beta \langle W_3 W_4 \rangle  { 1 \over  \partial^2 _\beta \log Z(\beta) } 
\ee
where $M$ is the mass of the black hole background, or $\beta$ its temperature, and $S(M)$ or $\log Z$ are its entropy or partition function.\footnote{
The correlator at finite $\beta$ is $\langle V V \rangle = \left[ { \beta \over \pi } \sin{ \pi u_{12} \over \beta } \right]^{ -2 \Delta } $.}
 Both expressions 
give the same answer, thanks to thermodynamic identities between entropy and mass.\footnote{(\ref{Eflu}) is valid for a general spherically symmetric
reduction of general relativity to two dimensions.} If one expands as $u_{12} \to 0$ we get a leading term going like $u_{12}^2$ which one would identify with an operator of dimension two. 
In this case this is the Schwarzian itself, which is also the energy and it is conserved \nref{ADM}. Its two point functions are constant.\footnote{
Note that this is different than in 1+1 dimensions, where the stress tensor correlators go like $1/z^4$.}

It is also interesting to evaluate the correlator in the other ordering 
$u_4 < u_2  < u_3 < u_1  $. We get  
\bea
  { \langle V_1 W_3 V_2 W_4 \rangle_{\rm grav} \over \langle V_1 V_2 \rangle \langle W_3 W_4 \rangle } &=& {   \Delta^2 \over 2 \pi C }
  \left[ 
  % \frac{\Delta}{\sin^{2\Delta}u_{12}}\frac{\Delta}{\sin^{2\Delta}u_{34}}
  %\left( 4-\frac{2u_{12}+2\pi}{\tan(\frac{u_{12}}{2})}+\frac{2\pi\cos(\frac{u_1+u_2}{2})}{\sin({u_{12}\over 2})} \right). 
\left(-2 +{u_{12}\over \tan {u_{12}\over 2}}\right)\left(-2+{{u_{34}}\over \tan {u_{34}\over 2}}\right) +
\right. 
\cr
&+ & \left. { 2 \pi [  \sin( { u_1 -u_2 + u_3 - u_4 \over 2 } ) -\sin( {u_1 + u_2 - u_3 - u_4 \over 2 } ) ] \over\sin {u_{12}\over 2}\sin{u_{34}\over 2} }
 + {2\pi u_{23}\over \tan {u_{12}\over 2}\tan{u_{34}\over 2}} \right] \la{Crossed}
\eea
This expression interpolates between \nref{Separated} when $u_3=u_2$ and an expression like \nref{Separated}, but with $u_{34} \to - 2\pi + u_{34}$, when
$u_3 =u_1$. 
Note that now the answer depends on the overall separation of the two pairs. This dependence, which involves the second
sine term in the numerator as well as the $u_{23}$ factor, looks like we are exciting the various zero modes of the Schwarzian action, including the exponential ones. It is interesting to continue \nref{Crossed}   to Lorentzian time and into the chaos region which involves the 
correlator in the out of time order form 
\be \la{ChaosRegime} 
\langle V( a) W_3(b+\ul) V(0) W(\ul) \rangle \sim   {  \beta \Delta^2 \over  C  }  e^{ 2 \pi \ul \over \beta }  ~,~~~~~ { \beta \over 2 \pi} 
 \ll \ul \ll { \beta \over 2 \pi }  \log{  C \over \beta } 
\ee
where $a , b \sim \beta$. Here we restored the temperature dependence in \nref{Crossed}  by multiplying by an overall a factor of $ { \beta \over 2 \pi }$
and sending $u_i \to { 2 \pi \over \beta } u_i$. 

We can also connect \nref{ChaosRegime} to a scattering process. It is peculiar that in this setup the two particles do not
scatter since they behave like free fields on a fixed $AdS_2$ background. On the other hand, they create a dilaton profile which 
gives rise to a non-trivial interaction once we relate the $AdS_2$ time to the boundary time. The net result is the same 
as what is usually produced  by the scattering of shock waves, see appendix \ref{ShockWaves}. Here we see that the gravitational 
effects are very delocalized, we can remove them from the bulk and take them into account in terms of the boundary degree of freedom $t(u)$.

\subsection{Loop corrections } 

We can use the Schwarzian action as a full quantum theory and we can compute loop corrections. 
The simplest example corresponds to the one loop correction to the free energy \cite{PolchStreich}. This arises from computing the functional determinant 
of the quadratic operator in \nref{Lora}. This was done in detail in \cite{Maldacena:2016hyu} 
   and we will not repeat the details. 
The important point is simply that it gives a temperature dependent correction to the free energy going like 
\be\label{foneloop}
\log Z |_{\rm one ~loop} = - { 3 \over 2} \log \left( { \beta \over C } \right)  
\ee
This is a correction to the leading classical expression \nref{SchS}. The determinants of all matter fields in $AdS_2$ are conformally invariant
and should not give rise to a temperature dependent contribution, but they can and do contribute to the extremal entropy \cite{Sen:2011ba,Banerjee:2011jp}. 
The correction (\ref{foneloop}) is such that there is no logarithmic correction to the entropy as a function of energy. This is good, since there are cases, such as
BPS black holes where do not expect corrections that diverge in the IR. 

As a second example we can consider a loop correction to the two point function. We expand the reparameterization
\be
{ 1 \over t_{12}^{2 \Delta } } \to  { ( 1+ \varepsilon'_1)^\Delta (1 + \varepsilon'_2 )^\Delta \over (\sin { u_{12} + \varepsilon_1-\varepsilon_2 \over 2 } )^{2\Delta} } 
\ee
to quadratic order in the $\varepsilon$, and then contract using the propagator (\ref{propag})
\bea
{ \langle V_1 V_2 \rangle_{\rm one~ loop} 
\over 
 \langle V_1 V_2 \rangle_{\rm tree} } &=&   \Delta \left\langle  { ( \varepsilon_1 - \varepsilon_2)^2 \over 4 \sin^2{ u \over 2 }} - { 1 \over 2 } ({\varepsilon'}^{\, 2}_1 + { \varepsilon'}_2^{\, 2} ) \right\rangle  +    
 { \Delta^2\over 2}  \left\langle \left( 
 \varepsilon'_1 + \varepsilon'_2 - {(  \varepsilon_1 -\varepsilon_2) \over \tan { u  \over 2 } }  \right)^2  
\right\rangle \notag
\\ \la{OLC}
&=& { 1 \over 2 \pi C } \left[ \Delta  { ( u^2 - 2 \pi u +2 - 2 \cos u+ 2(  \pi -  u ) \sin u ) \over 4 \sin^2{ u \over 2} }  
+\right.
\cr
&+& \left. \frac{\Delta^2 }{2} \left( -2 + { u \over \tan { u \over 2} } \right) \left( -2 + { (u - 2 \pi)  \over \tan { u \over 2} } \right)\right] ,~~ u = u_1 -u_2 >0
\eea 
It is interesting to continue these formulas to Lorentzian signature $ u \to i \ul $  and then expand them for large Lorentzian times. 
The largest term, which goes at $\ul^{\, 2}$ for large lorentzian time can arise from energy fluctuations in a manner analogous to 
\nref{Eflu}. The tree level correlator includes a quasinormal decay as $e^{ -  \Delta \taul} \sim e^{ -  \Delta 2 \pi \ul/\beta }$. But the energy fluctuations cause a temperature
fluctuation which would then lead to a correction for the ratio of the one loop to tree level as 
\be
   \frac{\Delta^2}{2} {4  \pi ^2\ul^2 \over \beta^4} { 1 \over \partial_\beta^2 \log Z}  =  { \Delta^2   \ul^2 \over 2\beta C }
\ee 
which agrees with the $\ul^2$ piece from \nref{OLC}.\footnote{ In fact, (\ref{OLC}) can be written exactly as $
\langle V_1 V_2 \rangle_{\rm one~ loop} 
= \frac{1}{2\partial_\beta^2\log Z}\partial_\beta \left\{\left[\partial_\beta \langle V_1 V_2\rangle_{tree}\right]_{u\rightarrow \beta-u}\right\}.$
The derivatives seem to be a way of varying the temperature in a way that maintains periodicity.}

\section{ Lorentzian picture and the SL(2) symmetry } 

\subsection{SL(2) symmetry of the Schwarzian action} 

We have seen that gravitational effects are summarized by the Schwarzian action \nref{SchAction}. 
This action seems problematic when viewed as an action in Lorentzian signature since it involves higher
derivative terms. These usually lead to ghosts. We can see this more  explicitly by starting with the 
Lorentzian action for small fluctuations 
\be \la{Lora}
i I_L =-i C \int d \ul \, \Sch(\tl, \ul)=i { C \over 2}  \int d \ul  ( { \varepsilon''}^{\, 2} + { \varepsilon' }^{\, 2} )~,~~~~~~~~{\rm for} ~~ \beta = 2 \pi ~,~~~C \equiv  { \bar \phi_r \over 8 \pi G} 
\ee
It is possible to rewrite this higher derivative action in terms of a two derivative action for two fields by introducing a new field $\eta$
\bea \la{KinTeg}
&&\int d\hat u [ \varepsilon''^{\, 2 } + {\varepsilon'}^{\, 2 } ] \to \int d\hat u 
\left[   \eta ( \varepsilon'' - { \varepsilon \over 2 }) - {\eta^2 \over 4 } - { \varepsilon^2 \over 4 }
\right]  =
\int d\ul  [  - r'^2 -r^2 + q'^{\, 2 } ]
\cr
&& \varepsilon = r + q ~,~~~\eta = r - q
\eea
Integrating out $\eta$ we get  $\eta = 2  \varepsilon'' - \varepsilon $ and recover the original action,
\nref{Lora}.  We can also use this expression for $\eta$ 
 to express 
$r$ and $q$ in terms of $\varepsilon$ which gives
\be
r = \varepsilon'' ~,~~~~~~~~~~ q = -\varepsilon'' + \varepsilon \label{qandr}
\ee
The full set of solutions of the original Lorentzian action \nref{Lora} is given by 
\be
\varepsilon = ( \alpha e^\ul + \beta e^{-\ul }  )  + ( \gamma \ul  + \delta )  \la{SolG}
\ee
We see that the first parenthesis corresponds to the ghost like mode $r$ and  the last one to the mode $q$. 
Note that in Euclidean space we started out with a full function worth of nearly zero modes, but in the Lorentzian theory these only give
rise to the two degrees of freedom $r$ and $q$. 

Should we be worried by the appearance of the ghost like mode that has a negative sign in its kinetic term in \nref{KinTeg}? Should we view the exponentially growing solutions in \nref{SolG} as an instability?
To answer these questions we need to recall that the original metric had an unbroken $SL(2)$ symmetry. 
Such $SL(2)$ diffeomorphisms do not generate
a new cutout geometry.   Thus, we should not include them in our integration over Pseudo-Goldstone modes.
 One way to remove them is to treat such diffeomorphisms as a gauge symmetry. 
More precisely, the Schwarzian action has an $SL(2)$ global symmetry. This global symmetry has its associated conserved charges 
\bea \la{SL2ch}
Q^{-} &=&C   \left[ { \tl''' \over \tl'^2 } - { \tl''^{\,2} \over \tl'^3 } \right]  =C e^{-\taul } \left[ { \taul''' \over \taul'^{\, 2} } -
 { \taul''^{\,2} \over \taul'^{\, 3} } +{\taul'' \over \taul' } \right]  ~,~~~~ 
\cr
Q^0 &=&  C \left[{\tl'''\tl \over \tl'^2 } - { \tl \tl''^{\,2} \over \tl'^3} - { \tl'' \over \tl' }   \right]= C   \left[  { \taul''' \over \taul'^{\, 2} } -
 { \taul''^{\,2} \over \taul'^{\, 3} } - {  \taul' } \right] 
\cr
Q^+ &= &C  \left[ { \tl''' \tl^2 \over\tl'^2 } - { \tl''^{\,2} \tl^2 \over \tl'^3 } - { 2 \tl \tl'' \over \tl'} + 2 \tl'   \right]=
C  e^{ \taul } \left[ { \taul''' \over \taul'^{\, 2} } -
 { \taul''^{\,2} \over \taul'^{\, 3} } - {\taul'' \over \taul' } \right] 
\eea
 where we also wrote it after setting $\tl = e^\taul $, which is appropriate for Lorentzian finite temperature computations. 
Treating them as  gauge symmetry amounts to saying that the full state should be invariant under these symmetries.  
However, we see that a solution with nonzero $ \taul'$ cannot have zero charges! Recall, though, that in the bulk this $SL(2)$ symmetry acts on the full $AdS_2$ spacetime.
 This means that it is a symmetry of the thermofield double. In the quantum mechanical description of the thermofield double we
have two sides and the charges are equal and opposite on the two sides. $Q^a_L = - Q^a_R$, so that the total charge can be zero. Therefore, purely on one side the charges can be anything. We can view the charges
$Q^a$ as proportional to the vector $Z^a$ in \nref{zexpr} that determines the location of the 
bifurcation point.  The SL(2) transformations move this point in $AdS_2$. This motion has no
physical consequence because the location of the boundary is determined by the value of the dilaton and
thus the boundary curve moves together with the bifucation point as we perform an SL(2) transformation.

 For the simplest solution $\taul = { 2 \pi \over \beta } u $, the charges are 
 \be
 Q^\pm =0 ~,~~~~~~~~~~Q^0 = -C  { 2 \pi \over \beta }    \la{ValQz}
 \ee
 This value of $Q^0$  (when $Q^\pm =0$) can be viewed as (minus) the near extremal 
entropy of the black hole. More precisely $S = -2 \pi Q$. As Wald has pointed out 
 \cite{Iyer:1994ys}, we can view black hole entropy as a Noether 
charge associated to the translation generated by  the horizon generating Killing vector. 
  
There is an additional conserved quantity of the Schwarzian action, which is simply associated to $u$ time translations. This is   the Hamiltonian discussed
in \nref{ADM}.  
%Note that $H$ should be viewed as the total energy of the system, it is the ADM energy of the system, reflecting the total energy content of the 
%state, accounting for possible matter inside. 
It is interesting to note the relation 
\be
 H = { 1 \over 2  C}  \left[ - Q^+ Q^- + (Q^0)^2 \right] \la{HandQ}
\ee
between the energy and the charges. Here the $Q$ are the charges of only the $t$ field on one side, as in \nref{SL2ch}. 
%Notice that the simplest  solution $\tau ={ 2 \pi \over \beta }  u $ breaks both $\tau$ translations as well as $u$ translations
%, keeping a linear combination unbroken. 
%This implies that $M = { 2\pi \over \beta } Q^0$ for the black holes, as we can check from \nref{ValQz} and \nref{HandQ}.  
%CHECK, FIX. 

It is also interesting to evalute the charges and the Hamiltonian for a first order 
 perturbation around the thermal solution, $\taul = u + \varepsilon(u)$ 
\bea
&& Q^\pm \sim  C  e^{ \pm u } \left[ \varepsilon''' \mp \varepsilon''\right] ~,~~~Q^0 \sim   C \left[ -1 + \varepsilon''' - \varepsilon' \right]
\cr
&& 
H \sim C  \left[\half - (\varepsilon''' - \varepsilon' ) \right] \label{Hexp}
\eea 
With these expressions we see that the zero mode $\varepsilon = e^{u}$ only contributes to $Q^-$ and $\varepsilon = e^{-u}$ only to $Q^+$.  
Two point functions of $H$ are constant, as expected for a conserved quantity.\footnote{Two point functions of the $Q^a$, such as 
$\langle Q^a(u)  Q^b(0) \rangle $ are {\it not} constant, despite their classical conservation law. This is due to the fact that we needed to 
break the SL(2) symmetry to compute the propagator for $\varepsilon$ \nref{propag}. This is not a problem because these are not gauge
invariant quantities. } Saying that we treat the SL(2) symmetry as a gauge symmetry implies that we 
are not free to excite these modes. These modes are excited in an amount that is set by the value of the charges. 

At this linear order in $\varepsilon$, we can also show that the Hamiltonian has the expected commutation relation with operators. From (\ref{Hexp}) and (\ref{qandr}) we have $H \sim C\left(\frac{1}{2} + q'\right)$. Assuming a canonical quantization of the non-ghost mode $q$, we conclude that $[H,\varepsilon] = C[q',q] = -i$ and $[H,\varepsilon'] = 0$. To evaluate the commutator with $V$, we include the reparameterization dressing $V \rightarrow (1 + \varepsilon')^\Delta V(u + \epsilon)$ and then expand to linear order in $\varepsilon$. This immediately gives $[H,V] = -iV'$.\\
%\ZY{[Or one can derive the commutation relation from propagator, for example express $[H,\epsilon] $ as integration of equation of motion from $0_-$ to $0_+$, and $[H,\epsilon']$ as difference of equation of motion as approach from $0_-$ vs $0_+$ ]}

\subsection{Adding matter} \label{matterSec}

If we have matter in $AdS_2$  then   the matter can also carry SL(2) charges. The total SL(2) charge is the sum of the matter one
plus the one carried by the field $\taul(\ul)$ that appears in the Schwarzian action. For massless matter we simply have
\be  \la{TotSL}
Q^a_T = Q^a(\taul) + q^a_M 
\ee
where $q^a_M$ are the standard charges associated to the $AdS_2$ isometries for the matter fields. The SL(2) gauge symmetry is saying that $Q_T$ will remain
constant as we add matter. This is compatible with the equations of motion \nref{MatEq}, and the fact that the SL(2) charges change  by a 
flux of energy; for massless matter we have simply $\partial_\ul ( Q^-,Q^0,Q^+) =  T_{\tl z} \tl'( 1,\tl,\tl^2 )$. When sources for massive fields are turned on, one has to add an extra stress tensor term to $q_M$ to define matter charges that satisfy the correct conservation conditions, see (\ref{improvedQ}).
% FIXSIGNS AND FACTORS. 
This SL(2) gauge symmetry implies that we cannot purely excite one of the ghost modes, we have to excite 
them together with some matter fields. 

The total energy is still given by the ADM expression \nref{ADM} and it is written purely in terms of the $\taul$ variable. In particular, 
\nref{HandQ} continues to be true where $Q^a$ in \nref{HandQ} are the SL(2) charges of the $\tau$ system only, they are {\it not} the total 
SL(2) charge appearing in \nref{TotSL}. So we see that the matter inside $AdS_2$ only carries SL(2) charge and their contribution to the
mass only appears through the SL(2) constraints that relate these charges to the SL(2) charge of the $\taul$ variable.

Suppose that we start from the thermofield double state and then we add matter on the right part. 
Then from the right part point of view we can view the charges and masses of the 
left part of the thermofield double as being carried at the horizon. Then the condition that the total charge vanishes becomes simply the condition that 
\be
 Q^a_h   = Q^a + q^a_M = Q^a_R  \la{Chaco}
 \ee
 where $Q^a_R$ are the total charges of the right system, and $q_a^M$ are the $SL(2)$ charge of possible matter falling into the black hole. 
 And $Q^a_h = - Q_L^a$ is the value of the charge at the bifurcation point, and also equal to minus the charges of the left side.
  Here we assumed that there is no matter on the left side of the spacetime. 
 We have seen that for the simple solution $\tau = u $ the charge $Q^0< 0$, and it is related to the 
energy \nref{HandQ}. 
 On the other hand, with the same conventions, the matter charge $q^0_M$ of a matter particle
would be {\it positive}. That makes \nref{Chaco} compatible with the energy conservation 
condition for small fluctuations which says that the mass of the black hole plus the energy of matter
should the be same as the energy measured at the boundary.

 When we throw matter into the black hole, the values of these charges change, but always in agreement with the conservation law. As we send in matter from the boundary, its additional mass is immediately 
recorded in the new value for the on shell Schwarzian. The SL(2) charges of the matter, together with the 
boundary charges, are  constrained to add to the same value that the boundary system had before we threw in 
the matter.  For an initial configuration with $Q^\pm =0$, the changes of the $Q^0$ charges 
 demanded by \nref{Chaco} can be viewed as a consequence of the first law, once we remember that
$Q^0$ is related to the entropy.

%In particular, suppose that 
% we have a given initial black hole with dilaton specified by $\Phi = Z. Y$, then the vector $Z$ is proportional to the three charges in 
%\nref{SL2ch}, $\vec Z \propto \vec Q $. Porforming SL(2) transformations amounts to changing the
%orientation of $Z$. The direction of $Z$ selects a point in $AdS_2$ which is the bifurcation surface of 
%the black hole solution. The SL(2) transformations move this point in $AdS_2$. This motion has no
%physical consequence because the location of the boundary is determined by the value of the dilaton and
%thus the boundary curve moves together with the bifucation point as we perform an SL(2) transformation. 

Let us add a classical massive particle following a geodesic in the background $AdS_2$ spacetime. The equations for this geodesic are given by 
$A. Y =0$ in embedding coordinates. If we choose $A^2 = \pm m^2$, then $\vec A$ is also proportional to the $SL(2)$ charges. 
Then the dilaton on the other side of the geodesic is given by $\Phi = (Z + A).Y$, where $Z+A$ reflect the $SL(2)$ charges on the other side of the geodesic.

 An important point to note is that the Schwarzian action breaks conformal symmetry for the $u$ time, so that general $SL(2)$ transformation of $u$, such as 
 $u \to ( a u + b)/(c u + d)$ is {\it not } a symmetry. (Only $u \to u + $constant is a symmetry.) Nevertheless the $SL(2)$ charges acting on $t$ are 
 still conserved, since they are gauge symmetries. In other words, we should not confuse the SL(2) symmetry acting on $t$, which is gauged and thus
 unbroken, with the SL(2) symmetry acting on $u$ which is not gauged, and it is broken by the Schwarzian action. 
 
 These charges are analogous to the edge modes of the electromagnetic field discussed in \cite{Donnelly:2014fua,Harlow:2015lma}, or the ``center'' in \cite{Casini:2013rba}, or horizon symmetries in 
\cite{Hawking:2016msc}.
 
 It is likely that there is a more elegant way to think about these SL(2) charges using the SL(2) gauge theory formulation of \nref{TJ} 
 \cite{Fukuyama:1985gg,Chamseddine:1989yz}.

\section{Higher orders in the chaos region}
The order $G\sim 1/C$ term in the four point function (\ref{ChaosRegime}) is exponentially growing in the time separation $\ul$ of the $V$ and $W$ operators. However, we actually expect the full correlator to become small at large $\ul$. This is due to higher order effects in powers of $1/C$. As an application of the Schwarzian action, we will show how to sum powers of $(e^{\ul}/C)$, which will be enough to capture the late-time decay. More precisely, we work in a limit $C\rightarrow \infty$, $\ul \rightarrow \infty$ with $e^\ul/C$ fixed. 

\subsection{Bulk inspiration }

  The procedure is equivalent to a bulk analysis where the tree-level amplitude is upgraded to an eikonal $\mathcal{S}$ matrix. We will briefly review this analysis (see \cite{Shenker:2014cwa}, using \cite{'tHooft:1987rb,Verlinde:1991iu,Kabat:1992tb}) which is very simple in $AdS_2$. The $V$ and $W$ operators are represented by bulk quanta with momenta $p_-$ and $q_+$, respectively. To capture all powers of $e^{\ul}/C$, one can replace the bulk metric by two effective ``shock wave'' modes. These are parameterized by shifts $X^+$ and $X^-$ on the future and past horizons of the black hole. The bulk metric perturbations and stress tensor are 
\be
h_{++} = 4\delta(x^+)X^-, \hspace{20pt} h_{--} = 4\delta(x^-)X^+, \hspace{20pt} T_{--} = -p_- \delta(x^-), \hspace{20pt} T_{++} = -q_+ \delta(x^+).
\ee
Here we are using Kruskal coordinates $x^+,x^-$ to describe the region near the horizon, see (\ref{shockm}). The bulk action $I_{JT} + S_M$ (\ref{TwoDAction}) for these quantities to quadratic order is
\be\label{bulkaction}
I_L = - 2CX^+ X^- - X^+q_+- X^-p_-.
\ee
The integral over $X^+,X^-$ of $e^{iI_L}$ gives the scattering matrix $\mathcal{S} = e^{ip_-q_+/2C}$. The four point function is an in-out overlap with this $\mathcal{S}$ matrix. The essential feature is that at late time the product $p_-q_+$ is large, and the $\mathcal{S}$ matrix implements a large translation of the wave packets, making the overlap small.
% \be
% \langle V_1W_3V_2W_4\rangle = \int dp^+dp^- \psi_1(p^+)\psi_2(p^+)\psi_3(p^-)\psi_4(p^-)\mathcal{S}(p^+,p^-)
% \ee

\subsection{Full resummation from the Schwarzian action } 
We will now do the calculation for real, using the boundary formulation of the theory. The metric is always exactly $AdS_2$; the only variable is the reparameterization $t(u)$. We have to identify $X^+,X^-$ with certain modes of $t(u)$ and then evaluate the Schwarzian action and the coupling to $V,W$. The main subtlety is that for the out-of-time-order correlator, we have to think about the function $t(u)$ on a folded time contour. 

We start with a single shock, so just $X^+$ is nonzero. The bulk solution consists of two black holes glued together with a shift along the horizon. In terms of $t(u)$, we glue two solutions together at $\ul = \infty$ with an SL(2) transformation. In an SL(2) frame where the original black hole solution is $\tl = e^{\ul}$, the transformation is a simple translation, and the shock wave solution is
\be
\tl = e^{\ul} \hspace{20pt} (\text{sheet 1}), \hspace{40pt} \hat{t} = e^{\ul} + X^+ \hspace{20pt}(\text{sheet 2}).\label{X+simple}
\ee
Notice that these can be glued at $\ul = \infty$. We will find it more convenient to work in a different SL(2) frame, where the original solution is $\tl = \tanh(\frac{\ul}{2})$, see figure \ref{CoordFig}. Then the one-shock solution is
\be
\tl = x \hspace{20pt} (\text{sheet 1}), \hspace{30pt}
\tl = x +\frac{(1-x)^2X^+}{2 + (1-x)X^+} \hspace{20pt} (\text{sheet 2})\label{X+}~,~~~~x\equiv \tanh(\frac{\ul}{2}).
\ee
This new SL(2) frame makes it clear that if $X^+$ is small then we have a small perturbation to $\tl$ for all values of $\ul$.

\begin{figure}[ht]
\begin{center}
\includegraphics[scale = 0.65]{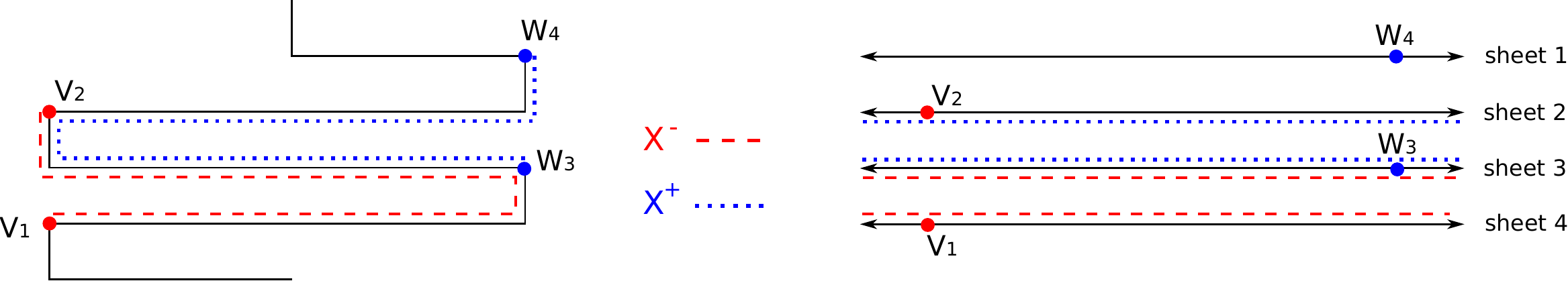}
\caption{The black contour at left is the minimal time contour needed to compute $\langle V_1W_3V_2W_4\rangle$. Horizontal is real time, vertical is imaginary time, the ends are identified. It is convenient to extend the folds to infinity, as shown at right. The blue and red indicate which sheets have the $X^-$ and $X^+$ perturbations turned on.}\label{foldFig}
\end{center}
\end{figure}
To compute the $\langle V_1 W_3 V_2 W_4\rangle$ correlator, we need a contour with four folds, as in figure \ref{foldFig}. We would like to superpose the (\ref{X+}) solution and a similar expression for $X^-$ in the right way to capture the important part of the functional 
integral over $t(u)$. We can guess the answer based on the bulk picture, or from the discussion of SL(2) charges in the previous section, which suggests that each pair of operators is associated to a relative SL(2) transformation between the portion of the contour inside and outside the pair. So we consider the configuration
\be\label{reparamans}
\tl = x + \frac{(1-x)^2X^+}{2 + (1-x)X^+}\theta(2,3)- \frac{(1+x)^2 X^-}{2 +(1+x)X^-}\theta(3,4), \hspace{20pt} x \equiv \tanh(\frac{\ul}{2}).
\ee
The $\theta$ symbols are defined to be equal to one on the sheets indicated and zero elsewhere, see figure \ref{foldFig}. This is not a solution to the equations of motion, it is an off-shell configuration of $t(u)$. The idea is that by integrating over $X^+,X^-$, we are capturing the part of the integral over $t(u)$ that gives powers of $e^{\ul}/C$. The entire dependence of the action (\ref{SchAction}) on $X^+,X^-$ comes from the fold where both terms are nonzero. The product $X^+X^-$ is small, of order $1/C$ in the limit we are taking, so it is enough to compute the action to quadratic order:
\be\label{schac12}
iI_L \supset - i C\int_{\text{sheet }3} d \ul \  Sch(\tl,\ul) = -2iCX^-X^+ + (X\text{-independent}).
\ee

To compute the four point function, we also have to consider the reparameterized two-point functions. If the $V$ operators are at early time, then only the $X^+$ part of the reparameterization is important. It acts on sheets two and three, and therefore affects only $V_2$ (see figure \ref{foldFig}). Similarly, for the $W$ operators we only have to consider the $X^-$ part acting on $W_3$:
\begin{align}
G_V(\ul_1,\ul_2) = \left[\frac{-\tl'(\ul_1)\tl'(\ul_2)}{(\tl(\ul_1) - \tl(\ul_2))^2}\right]^\Delta&\approx \left[\frac{-i}{2\sinh\frac{\ul_{12}}{2} - X^+e^{-(\ul_1+\ul_2)/2}}\right]^{2\Delta}\label{GV}\\
G_W(\ul_3,\ul_4) = \left[\frac{-\tl'(\ul_3)\tl'(\ul_4)}{(\tl(\ul_3) - \tl(\ul_4))^2}\right]^\Delta&\approx \left[\frac{-i}{2\sinh\frac{\ul_{34}}{2} - X^-e^{(\ul_3+\ul_4)/2}}\right]^{2\Delta}\label{GW}%\\
%&=\frac{1}{\Gamma(2\Delta)}\int_0^\infty \frac{dp}{p}  \Psi_1(p)\Psi_2(p)e^{iX^+p}.
\end{align}
Now, to compute the four point function, we simply integrate these expressions over $X^\pm$ with the weighting given by (\ref{schac12}). Up to measure factors, we have
\be\label{insertstep}
\langle V_1W_3V_2W_4\rangle \propto \int dX^+dX^- e^{-2iCX^+X^-}G_V(\ul_1,\ul_2)G_W(\ul_3,\ul_4).
\ee
\begin{figure}[t]
\begin{center}
\includegraphics[scale = 0.7]{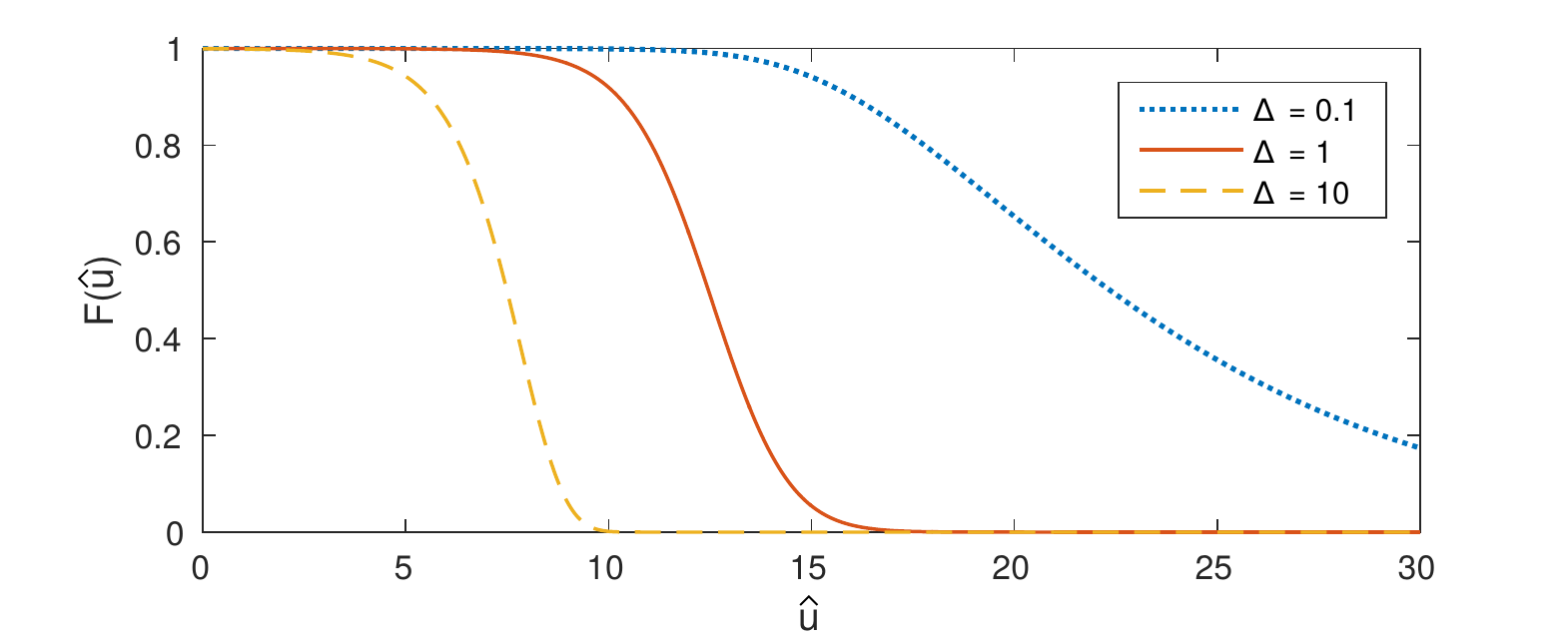}
\caption{The correlation function (\ref{resummed}) in the configuration (\ref{equalspacing}), with $8C = 10^6$. Scaling up $C$ simply translates all of the curves to the right without changing their shape. The initial descent from the plateau is characterized by the $e^{\frac{2\pi}{\beta}\ul}$ behavior. The final approach to zero is determined by quasinormal decay.}\label{plotFig}
\end{center}
\end{figure}
One of the integrals can be done simply, and the second can be expressed using the confluent hypergeometric function $U(a,1,x) = \Gamma(a)^{-1}\int_0^\infty ds e^{-sx} \frac{s^{a-1}}{(1+s)^{a}}$. The answer is
\be\label{resummed}
\frac{\langle V_1 W_3V_2W_4\rangle}{\langle V_1 V_2\rangle\langle W_3 W_4\rangle} = \frac{U(2\Delta,1,\frac{1}{z})}{z^{2\Delta}}, \hspace{20pt} z = \frac{i}{8C}\frac{e^{(\ul_3+\ul_4-\ul_1-\ul_2)/2}}{\sinh\frac{\ul_{12}}{2}\sinh\frac{\ul_{34}}{2}}.
\ee
The real part of $z$ is positive for the ordering of operators we have assumed. A simple configuration to keep in mind is the one where the $V,W$ operators are equally spaced around the Euclidean circle, for example
\be
\ul_1 = -\frac{\ul}{2} - i\pi, \hspace{15pt} \ul_2 = -\frac{\ul}{2}, \hspace{15pt} \ul_3 = \frac{\ul}{2}-i\frac{\pi}{2}, \hspace{15pt} \ul_4 =\frac{\ul}{2}+ i\frac{\pi}{2} \hspace{15pt} \implies \hspace{15pt} z = \frac{e^\ul}{8C}.\label{equalspacing}
\ee
Here $\ul$ is the separation of the early $V$ operators and the late $W$ operators. Notice that the $z$ variable is real and positive in this type of configuration. We give a plot of (\ref{resummed}) in figure \ref{plotFig}. For early and late times, we have the limiting behaviors
\be
\frac{U(2\Delta,1,\frac{1}{z})}{z^{2\Delta}}  \approx   1 - 4\Delta^2 z \hspace{20pt} (z\ll 1), \hspace{40pt} \frac{U(2\Delta,1,\frac{1}{z})}{z^{2\Delta}}\approx \frac{\log z}{\Gamma(2\Delta)z^{2\Delta}} \hspace{20pt} (z\gg 1).
\ee
The small $z$ expression reproduces the initial exponential growth of the connected correlator (\ref{ChaosRegime}). The large $z$ behavior gives exponential decay of the full correlator at late time, 
where it is dominated by the decay of the quasinormal modes. 

\subsection{Role of the SL(2) charges}
To understand the charges of the matter, it is helpful represent (\ref{GV}) and (\ref{GW}) in a basis that diagonalizes certain SL(2) generators. It is convenient to return to the SL(2) frame where the background solution is $\tl = e^{\ul}$. One can write
\begin{align}
G_V(\ul_1,\ul_2) &= \frac{1}{\Gamma(2\Delta)}\int_{-\infty}^0 \frac{dq_+}{-q_+}\Psi_1(q_+)\Psi_2(q_+)e^{-iX^+q_+}\label{GW3}\\
G_W(\ul_3,\ul_3) &= \frac{1}{\Gamma(2\Delta)}\int_{-\infty}^0 \frac{dp_-}{-p_-}\Psi_3(p_-)\Psi_4(p_-)e^{-iX^-p_-}\label{GV3}
\end{align}
where the wave functions are given by
\be
\Psi_j = x_j^\Delta e^{-x_j}, \hspace{15pt} x_1 = -iq_+e^{\ul_1}, \hspace{15pt} x_2 = iq_+e^{\ul_2}, \hspace{15pt} x_3 = ip_-e^{-\ul_3}, \hspace{15pt} x_4 = -ip_-e^{-\ul_4}.
\ee
Equations (\ref{GW3}) and (\ref{GV3}) decompose the bilocals into pieces with definite charge, e.g. $q_M^- = -q_+$. Based on the discussion in section \ref{matterSec}, we expect these charges to be related to the charges of the shocks in $t(u)$. For example, when we pass the $V$ operator, the $X^-$ shock turns on, and one can check from (\ref{SL2ch}) that the charge changes by $Q^-_{after} - Q^-_{before} = -2CX^-$. This gets related to $q_+$ as follows: when we insert (\ref{GW3}) and (\ref{GV3}) into (\ref{insertstep}) and integrate over $X^+$, we will get a delta function setting $-2CX^--q_+ = 0$, so that
\be
Q^-_{after} - Q^-_{before} = -q_M^-.
\ee
This means that the total charge $Q_T$ indeed remains constant.

The charges give a new perspective on the exponential growth of the four point function. Acting with $V$ at an early time changes $Q^-$ of the $\taul$ sector. Because the charge is conserved, and because of the explicit factor of $e^{-\taul}$ in (\ref{SL2ch}), this has an exponentially growing effect on $\taul(u)$ as we move toward the future. We can always make an SL(2) gauge transformation to remove this effect on either the portion of the contour before $V$ or the portion after, but not both. In the out-of-time-order four point function we have $W$ operators probing both sides, so the exponential effect is physical.

\section{Discussion } 

We have pointed out how the asymptotic  symmetries of $AdS_2$ can be used to determine many aspects
of the gravitational dynamics of nearly $AdS_2$ spacetimes, or $NAdS_2$.  
The essential feature is the emergence of a reparametrization symmetry which is both spontaneously and explicitly broken. 
The corresponding pseudo-Goldstone bosons are described by a reparametrization $t(u)$ that expresses $AdS$ time $t$  in terms of the
physical boundary time $u$. The explicit breaking leads to a Schwarzian action for $t(u)$ \nref{SchAction}. In addition, we also have a simple coupling
to bulk fields \nref{sourct}. These together give rise to several features of $NAdS_2$ or near extremal black holes. 
These include the computation of the near extremal free energy as well as several gravitational effects involving correlation functions. 
These include gravitational corrections to the four point function \nref{Separated} and \nref{ChaosRegime} as well as corrections to the 
two point function \nref{OLC}. 

For all these features it was important to assume that the Schwarzian action was the leading effect that breaks the reparametrization symmetry.
This is the case in many interesting physical situations. However, one can imagine cases involving $AdS_2$ spaces with particularly light fields, 
dual
 to operators with dimensions $1 < \Delta < 3/2$. In these cases, if these fields are excited, then we have larger irrelevant perturbations 
and the infrared dynamics is different. See appendix \ref{NonSch} for a detailed discussion. 

The Schwarzian action involves higher derivative terms, which raise ghost fears. The ghosts are made invisible by treating the SL(2) symmetry 
of the Schwarzian action as a gauge symmetry. This reflects the fact that the whole configuration, including the boundary, can be shifted around
in $AdS_2$ space with no physical consequence. This is distinct from the physical SL(2) symmetry acting on $u$ which is broken by the Schwarzian 
action. 
The ghost-like degrees of freedom lead to exponentially growing corrections in the out of time ordered configuration. 

Note that the Schwarzian action has the flavor of a hydrodynamical theory. Namely, it reproduces 
the thermodynamics of the system. The fact that the entropy is the conserved charge associated
to  the $\tau$ circle translations also resonates with recent discussions of a $U(1)_T$ symmetry in \cite{Haehl:2015foa}, see also 
\cite{Jensen:2016pah}. It is also important
to include both sides of the thermofield double to make sense of the SL(2) constraints. 
Now, this Schwarzian action goes beyond ordinary long distance hydrodynamics, because it is 
including modes whose time variation rate is comparable to the temperature. Such modes 
are crucial for reproducing the out of time order correlator in the chaos regime. 

Two dimensional black holes are a very useful testing ground for ideas for solving the information 
paradox.  Any general idea should work in this simplest context.
A important element seems to be a better understanding of the emergence  of the 
charges $Q^\pm$. These are the symmetries that allow us to move into the interior!. 
 These charges are analogous to the edge modes of the electromagnetic field discussed in \cite{Donnelly:2014fua,Harlow:2015lma}, or the ``center'' in \cite{Casini:2013rba}, or horizon symmetries in 
\cite{Hawking:2016msc} (see also \cite{Donnelly:2016auv})\footnote{  It was emphasized in \cite{Hawking:2016msc} that the number of charges is infinite in more than two dimensions.
In two dimensions we simply have a finite number of charges,   the SL(2) charges discussed here.   }.

 There are several other  questions remaining to be answered. 
 In two dimensions similar holomorphic reparametrizations give rise to a Virasoro algrebra with a central charge. 
 Here we have  mentioned neither the algebra nor the central charge. It would be nice to see whether and how it can be defined. 
 Several papers have discussed a central charge for $AdS_2$, including 
 \cite{Strominger:1998yg,Cadoni:1999ja,NavarroSalas:1999up,Hartman:2008dq}, but we have not understood how they are connected to the 
 present discussion.

{\bf Acknowledgments } 

We thank A. Almheiri, D. Anninos, M. Dodelson, A.Kitaev and S.H. Shenker for discussions. 
J.M. is supported in part by U.S. Department of Energy grant
de-sc0009988.
 D.S. is supported by the Simons Foundation grant 385600.

\appendix 

\section{Massive fields in $AdS_2$ and their coupling to gravity}
\la{MassiveFields} 
In this appendix we study in some detail the effect that sources for massive fields have on the time-dependence of the Hamiltonian and the SL(2) charges. A subtlety is that the Schwarzian is not equal to the ADM Hamiltonian while such sources are turned on, it differs by a term involving $T_{zz}$. For the SL(2) charges, one has to add a similar term to the naive matter charges to get exact conservation.

We will start by considering free fields. We imagine we add classical sources at the boundary  by specifying the boundary 
conditions $\chi_r(u)$, see \nref{ChBdyC}. 
As we explained in section \nref{AddMat}, we can go from the effective action \nref{CorPoi} to \nref{sourct}. 
We can add this to the Schwarzian action \nref{SchAction} and then vary the resulting effective action for $t(u)$ to obtain 
a new classical equation
\bea 
C {    [ \Sch(t,u)]'  \over t' } &=& -{ 1 \over t'}  \left\{ \chi_r'(u)  {\cal O}(u)  + \partial_u \left[ ( \Delta -1)  \chi_r(u) {\cal O}(u) \right] \right\}
\la{EoMm}  \\
&& {\rm with}~~~ {\cal O}(u) \equiv   2 D\, t'(u)^{\Delta } \int du'{ t'(u')^\Delta \chi_r(u') \over (t(u) - t(u'))^{2\Delta } } \la{DefO}
\eea
where ${\cal O}$ can also  be interpreted as the classical expectation value of the operator dual to the souce $\chi_r(u)$. $\mathcal{O}$ and $\chi_r$ are related to the small $z$ behavior of the field by
\be\label{smallz}
\chi(z,t) = \left[\chi_r(u) - \frac{\epsilon^{2\Delta-1}\mathcal{O}(u)}{2\Delta-1}\right]\left(\frac{z}{t'(u)}\right)^{1-\Delta} + ...+ \frac{\mathcal{O}(u)}{2\Delta-1}\left(\frac{z}{t'(u)}\right)^\Delta+....
\ee
The explicit $\epsilon$ term is to ensure $\chi(\epsilon t'(u),t(u)) = \epsilon^{1-\Delta}\chi_r(u)$ so that (\ref{ChBdyC}) is satisfied.

We now want to relate (\ref{EoMm}) to the energy conservation condition. In the bulk, given any vector $\zeta^\mu$, we can 
construct a current $ (* j_\zeta )_\mu = \epsilon_{\mu ~}^\nu T_{\nu \delta } \zeta^\delta $, which is conserved when $\zeta$ is a Killing vector. 
In general the ADM mass $M$ is given by the first equality in \nref{ADM} as a function of the dilaton. We expect that its first 
derivative should give us the flux of energy into the system  
\be \la{massvar} 
\partial_u M = \frac{1}{8\pi G}  \partial_u (  t' \partial_z \phi - z' \partial_t \phi) =  (*j )_\mu (t' ,z')^\mu = \sqrt{h} T_{u n } 
% {\bf z' z' }T_{\bf zz } - {\bf \bar z' \bar z' } T_{\bar z \bar z} 
\ee
%CHECK SIGNS
%where ${\bf z} = t + i z $, ${\bf \bar z } = t - i z $. 
where $u$ is the coordinate along the boundary and $n$ is the normal direction. 
Here we have equated the flux with the energy corresponding locally to a direction tangent to the 
boundary curve. We also assumed $\partial_u \phi_b = 0$.
For a scalar field we see that 
 \be  \la{Massvar}
 \partial_u M = \sqrt{h} T_{u n} = \sqrt{h} \partial_u \chi \partial_n \chi= \partial_u \chi_r \left[ \epsilon^{1-2 \Delta } (\Delta -1 ) \chi_r - { \cal O } \right] 
 \ee
 where we used (\ref{smallz}). Note, that, as expected, energy is conserved as long as the sources are time independent. 
 The first term diverges as $\epsilon \to 0$ and can be cancelled by a counter term. Here we assumed that $ 1 < \Delta < 3/2$ in order to avoid
 further divergent terms.

 Comparing this with \nref{EoMm} we conclude that
 \be \la{SchM}
 M = C \Sch(t,u) + \epsilon^{1 - 2 \Delta } (\Delta-1) { \chi_r(u)^2 \over 2 } +   ( \Delta -1)  \chi_r(u) {\cal O}(u) 
 \ee
 
 In fact, we can compute the relation between the mass $M$ and the Schwarzian directly by using the definition of the ADM mass 
 \bea
 (8\pi G)M &\equiv & 
 %t' \partial_z \phi - z' \partial_t \phi + \phi \sqrt{h} =
  - \partial_n \phi  + \phi \sqrt{h} =
  t' \partial_z \phi - z' \partial_t \phi +  { \bar \phi_r \over \epsilon^2 } 
  = \la{ADMdef}
  \\
  &=& t' ( { \phi \over z} + \partial_z \phi ) - { z' \over z}  \partial_t (z \phi) - \left[  { t' \over z } - { 1 \over \epsilon } \right] {\bar \phi_r \over \epsilon } 
  \la{ThreeT}\eea
  where $\sqrt{h} =1/\epsilon$ is the boundary metric and $\phi = \bar \phi_r/\epsilon$ at the boundary \nref{phiequ}. 
  We have added and subtracted various terms. In the first term we use the $T_{zz}$ equation
  \be
  \partial_t^2 \phi - { 1 \over z^2 } \phi - { 1 \over z } \partial_z\phi =  8\pi G\, T_{zz}. 
  \ee
  In the third term in  \nref{ThreeT}  we expand the constant proper length condition as $ t'/z = { 1 \over \epsilon } - { 1\over 2} \epsilon { { t''}^2 \over {t'}^2 } $. 
  We also assume that 
  \be 
  \phi = { \phi_- \over z } + \phi_{sl}   
  \ee
  where $\phi_{sl}$ is less singular than $1/z$, so that $z\phi$ has a finite limit.
  Then we see that $\phi_- = t' \bar \phi_r $ to leading order. 
   We can also  convert $\partial_t$ into ${ 1 \over t' } \partial_u $ for the terms that are finite. 
  All these terms together then give 
  \be
M=  \frac{1}{8\pi G}\left[t' ( \partial_t^2 (z \phi) - 8\pi G z T_{zz} ) - { {t''}^2 \over {t'}^2 } \bar \phi_r + { 1 \over 2}  { {t''}^2 \over {t'}^2 } \bar \phi_r\right] = 
C\, \Sch(t,u) - t' z T_{zz}. \label{genphiS}
\ee
This derivation of the relation between the Schwarzian and the mass is valid also for a self interacting matter theory (that is not directly 
coupled to the dilaton in the lagrangian). 
Evaluating $zT_{zz}$ for a free field we obtain the extra terms in \nref{SchM}.
 
We can similarly consider the expressions for the SL(2) charges. We expect that the total SL(2) charges should be preserved {\it even with 
time dependent boundary conditions}. We first define a naive  matter SL(2) charge, $q^{(k)}_M$,  as the integral of $*j_\zeta$ over a spatial slice with 
$\zeta^\mu= ( k(t), \epsilon k'(t))$ near the boundary, with $k(t) = 1,~t,~t^2$, for each of the SL(2) generators. 
For a free field, the fluxes are then given as 
\be
\partial_u q_M^{(k)} =  -\sqrt{h}T_{n \mu} \zeta^\mu = { k \over t' } \chi_r'(u) {\cal O}(u) -  (\Delta -1) \left( { k \over t' } \right)' \chi_r { \cal O } 
+ \partial_u \left(  (1-\Delta) \epsilon^{ 1 - 2 \Delta } { \chi_r^2 k \over 2 t' }  \right)  
\ee
We now define a new matter charge that includes some extra terms of the form  
\bea
Q^{(k)}_M &=& q^{(k)}_M - k  z T_{zz} 
\cr
Q^{(k)}_M & = & \int * j_{\zeta }  +  (\Delta -1) \epsilon^{ 1 - 2 \Delta } { \chi_r^2 k \over2  t' } + (\Delta -1)  { k \over t' }   \chi_r { \cal O }  \label{improvedQ}
\eea
The extra terms are boundary terms that we can add in the definition of the charge. 
Then we see that the total SL(2) charges defined as 
\be
Q^{(k)}_T = Q^{(k)} + Q_M^{(k)}  ~,~~~~~~~~ \partial_u Q_T^{(k)} =0 
\ee
are conserved, once we use the equations of motion \nref{EoMm}. Here $Q^{(k)}$ are the charges constructed purely out of the $t(u)$ variable as
in \nref{SL2ch}, $(Q^-,Q^0,Q^+) = (Q^{(1)},Q^{(t)},Q^{(t^2)})$. They obey 
\be
\partial_u Q^{(k)} = C{ k  [ \Sch(t,u) ]' \over t' } ~,~~~~~~k =1,~t,~t^2.
\ee

\section{Gravitational shock wave scattering } 
\la{ShockWaves}
% Here we consider two pulses one created by $V(t)$ and one by $W(0)$ that scatter in two dimensional 
% gravity. 

% Discuss here the solutions. 

% What is important to point out is that the waves just go through each other in the usual AdS coordinates
% but that nevertheless there is an effect on the boundary time. 
% Thus, we seem to encode chaos in non interacting systems... 
% In other words, the growth of the commutator is only due to this boundary gravitons. 

Here we consider the scattering of two pulses in two dimensional gravity. This is simplest to discuss in a frame where one pulse is highly boosted, created by $V(-\ul)$ where $\ul$ large, and the other is unboosted, created by $V(0)$. As in higher dimensions, the scattering can then be described by studying the propagation of the probe $V$ particle on the background created by $W$ \cite{'tHooft:1987rb}.

\begin{figure}[h]
\begin{center}
\includegraphics[scale=1]{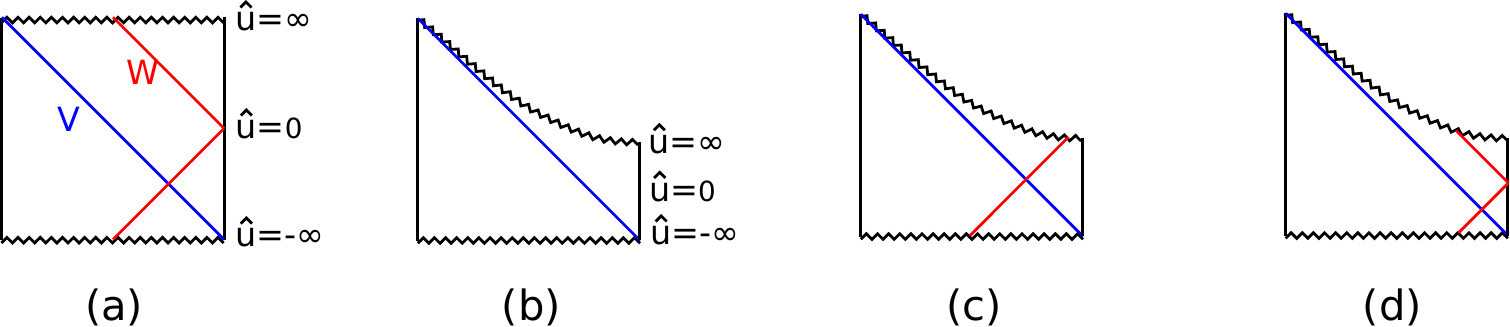}
\caption{In (a) we show the trajectories of the $V,W$ quanta without backreaction. In (b) we show the backreaction of the $V$ particle. This is still a piece of $AdS_2$, but it is a smaller piece. In (c) we add back $W$ in the arrangement appropriate for the operator ordering $V(-\ul)W(0)$. The trajectories are (almost) the same as (a) relative to the fixed $AdS_2$ coordinates of the diagram, but they change relative to the physical $\ul$ coordinate. In (d) we show the other ordering $W(0)V(-\ul)$. Now the red line touches the boundary at time $\ul = 0$. Although it is difficult to see in this frame, the $V$ line has moved down slightly, so that it no longer reaches the boundary.}
\label{appfig}
\end{center}
\end{figure}

In the theory (\ref{TwoDAction}), the metric is always exactly $AdS_2$, so there must be a set of coordinates in which this background is trivial, and the particles simply pass through each other, without detecting any local gravitational effect. If this is the case, how can there be any scattering at all? The answer is that these coordinates are related to the physical boundary coordinate $\ul$ in a nontrivial way \cite{Verlinde:1991iu,Almheiri:2014cka}. This is simplest to explain with a drawing, which we attempt in figure \ref{appfig}.

We can also relate this discussion to the standard shock wave picture. The backreaction of $V$ can be described by a metric
\be\label{shockm}
ds^2 = -\frac{4dx^+dx^-}{(1+x^+x^-)^2} + 4X^-\delta(x^+)(dx^+)^2
\ee
where $X^-$ is proportional to the large $p_+$ momentum of $V$. In these coordinates, the dilaton is a function of $x^+x^-$ only, and the physical time coordinate $\ul$ at the right boundary ($x^+x^- = -1, x^+>0$) is given by $e^\ul = x^+$. However, it is not manifest that they describe a piece of $AdS_2$. It is simple to check that we can rewrite (\ref{shockm}) as
\be
ds^2 = -\frac{4d\tilde{x}^+d\tilde{x}^-}{(1 + \tilde{x}^+\tilde{x}^-)^2}, \hspace{20pt} \tilde{x}^+ = \begin{cases} 
      x^+ & x^+<0 \\
      \frac{x^+}{1 + X^-x^+} & x^+>0,
   \end{cases}
    \hspace{20pt} \tilde{x}^- =  x^- - X^-\theta(x^+).
\ee
These coordinates make it clear that we have a piece of $AdS_2$. In fact, these are the coordinates of the fixed $AdS_2$ space on which the drawings in figure \ref{appfig} are represented. In these coordinates, the dilaton profile does not simply depend on $\tilde{x}^+\tilde{x}^-$. Their relationship to the time $\ul$ at the right boundary depends on $X^-$. At the boundary we have
\be
e^\ul = x^+ = \frac{\tilde{x}^+}{1 - X^-\tilde{x}^+}.
\ee

In summary, the particles simply pass through each other in the bulk, but nevertheless there is an effect on the boundary time. In this way, the gravity dual manages to encode chaos in non-interacting particles. 

%What is important to point out is that the waves just go through each other in the usual AdS coordinates but that nevertheless there is an effect on the boundary time. Thus, we seem to encode chaos in non interacting systems... In other words, the growth of the commutator is only due to this boundary gravitons. 

  \section{ Corrections to the matter two point functions}

It is natural to ask about the form of the leading correction to the matter two point functions due to futher couplings to 
the dilaton field such as 
\be
   \int d^2x \sqrt{g} \left[  ( \nabla  \chi )^2 + m^2 \chi^2  + \alpha  \phi \chi^2 \right] 
\ee
We can now consider the background value for $\phi = \phi_h \cosh \rho $ to find 
the leading correction to the thermal two point function. This can be found by noticing that the 
integral we need to do for the dilaton field
has the same form as the one expected for the insertion of a bulk to boundary propagator for a $\Delta =-1$ boundary operator. Thus the
correction to the correlator has the form $ \int du \langle O_{\Delta } (u_1) O_{\Delta }(u_2)  V_{-1}(u) \rangle$. Since the three point function
is fixed by conformal symmetry we find that  
\be
\langle O(1) O(2) \rangle = \left( {  \pi  \over \beta  \sin {\pi  \tau_{12} \over \beta } }\right)^{2 \Delta } 
\left[  1 + c_0 \alpha { \phi_r \over \beta } \left ( 2 +  \pi { (1    -  2 \tau_{12}/\beta)  \over \tan{ \pi \tau_{12} \over \beta } } \right) \right] 
\ee
where $c_0$ is a numerical constant. We see that this correction depends on a new parameter $\alpha$ 
that depends on the details of the theory. This has the same form as the 
corrections found in \cite{Maldacena:2016hyu} for the Sachdev-Ye-Kitaev model. 

\section{A case where the Schwarzian is not dominating} 
\la{NonSch}

Throughout this paper we have considered nearly $AdS_2$ situations where the Schwarzian is the leading irrelevant deformation. 
We now ask the question of whether this always happens or whether there are also situations where other corrections dominate. 

For simplicity we will focus only on systems that have a large $N$ expansion, or a weakly coupled gravity description. 
By assumption we have an IR fixed point, therefore we assume that there are no relevant operators turned on. 
We can consider the effects of turning on irrelevant single trace operators which correspond to changing the boundary values of 
massive bosonic fields in $AdS_2$. 

We will see that if we turn on an operator with 
\be \la{RangeD}
 1 < \Delta < 3/2
 \ee
  then its effects dominate over the ones due to the Schwarzian action and
the IR dynamics is different from the one described in this article. 

As before, we still have the zero modes in the IR parametrized by the field $t(u)$. But due to the presence of the irrelevant 
operator with dimension $\Delta$ 
we get an effective action given by \nref{sourct}
\be \la{Othop}
 -I_{eff}  = \lambda^2 \int du du' \left[ t'(u) t'(u') \over ( t(u) - t(u'))^2 \right]^\Delta 
 \ee
 here $\lambda$ is the coefficient of the operator in the action $ \int du \lambda O(u) $.  
 As a simple check that we obtain some effect that dominates over the Schwarzian, we consider the finite temperature 
 configuration with $t = \tan{ \pi u \over \beta }$ where we obtain a free energy of the form 
 \bea
 \log Z &=& \lambda^2 \beta \int_\epsilon^{\beta - \epsilon} du  \left[  \pi \over \beta \sin  { \pi u \over \beta } \right]^{2 \Delta } 
 \cr
 &=& \beta \lambda^2  { 1 \over \Delta -\half }  { 1 \over \epsilon^{ 2 \Delta -1} } + \lambda^2 \beta^{ 2 - 2 \Delta } { \pi^{ 2 \Delta -\half} \Gamma( \half - \Delta ) \over \Gamma(1 - \Delta ) } 
\eea
The first term is a UV divergence, but is proportional to $\beta$ so that it is a correction to the ground state energy. The second term is finite. 
We see that if $\Delta < 3/2$,  this second term dominates, for large $\beta$, over
 the Schwarzian answer \nref{SchS}, which goes as $\beta^{-1}$. 

Thus in the range \nref{RangeD} this operator gives the leading IR correction. The effective action for the reparametrizations \nref{Othop} is non-local. 
This can be checked more explicitly by setting $ t= u +\varepsilon(u)$ and expanding in fourier space. 
We end up with an action of the form $I_{eff} \propto  \int dp  |p|^{1 + 2 \Delta } \varepsilon(p) \varepsilon(-p) $.   
which indeed has a non-local form for  the range \nref{RangeD}. 

The Sachdev-Ye-Kitaev model \cite{Sachdev:1992fk,KitaevTalks}  has no operators in the range \nref{RangeD} \cite{KitaevTalks}, therefore the Schwarzian dominates in the IR. 

\section{ Lack of reparametrization symmetry in ``conformal'' quantum mechanics }

There are simple quantum mechanical theories that display an SL(2) conformal symmetry. 
An example is a lagrangian of the form 
\be \la{CQM}
 S = \int d t \left[  \left( { d X \over d t } \right)^2 - { \ell^2 \over X^2 } \right]
\ee
Under a transformation of the form 
\be 
 t \to t(\tilde t ) ~,~~~~~~~~~~ X(t(\tilde t) )=  (t')^{\frac{1}{2}} \tilde X(\tilde t ) 
\ee
\nref{CQM} changes to 
\be
S \to \int d \tilde t \left[ \left( { d \tilde X \over d \tilde t } \right)^2 - { \ell^2 \over \tilde X^2 }  - { 1 \over 2} 
\Sch( t, \tilde t ) \tilde X^2  \right]
\ee
We see that if $t(\tilde t)$ is an SL(2) transformation, then the action is invariant. However, if 
it is a more general reparametrization the action is not invariant.

\mciteSetMidEndSepPunct{}{\ifmciteBstWouldAddEndPunct.\else\fi}{\relax}
\bibliographystyle{utphys}
\bibliography{GravityReparametrizationsSubmitted.bib}{}

\end{document}